\documentclass[aps,prd,reprint,superscriptaddress,amsmath,amssymb,nofootinbib]{revtex4-2}
\usepackage[dvipsnames]{xcolor}
\usepackage{dcolumn}
\usepackage{graphicx}
\usepackage{hyperref}
\usepackage{multirow}
\usepackage{tabularx}
\hypersetup{colorlinks=true,linkcolor=blue}
\DeclareMathOperator\arctanh{arctanh}
\newcolumntype{Y}{>{\centering\arraybackslash}X}

\usepackage{scalerel}
\usepackage{tikz}
\usetikzlibrary{svg.path}
\definecolor{orcidlogocol}{HTML}{A6CE39}
\tikzset{orcidlogo/.pic={
    \fill[orcidlogocol] svg{M256,128c0,70.7-57.3,128-128,128C57.3,256,0,198.7,0,128C0,57.3,57.3,0,128,0C198.7,0,256,57.3,256,128z};
    \fill[white] svg{M86.3,186.2H70.9V79.1h15.4v48.4V186.2z}
    svg{M108.9,79.1h41.6c39.6,0,57,28.3,57,53.6c0,27.5-21.5,53.6-56.8,53.6h-41.8V79.1z M124.3,172.4h24.5c34.9,0,42.9-26.5,42.9-39.7c0-21.5-13.7-39.7-43.7-39.7h-23.7V172.4z}
    svg{M88.7,56.8c0,5.5-4.5,10.1-10.1,10.1c-5.6,0-10.1-4.6-10.1-10.1c0-5.6,4.5-10.1,10.1-10.1C84.2,46.7,88.7,51.3,88.7,56.8z};}}
\newcommand\orcidicon[1]{\href{https://orcid.org/#1}{\mbox{\scalerel*{
\begin{tikzpicture}[yscale=-1,transform shape]
\pic{orcidlogo};
\end{tikzpicture}
}{|}}}}

\begin{document}
\title{Testing a nonlinear solution of the Israel-Stewart theory}

\author{Miguel Cruz\orcidicon{0000-0003-3826-1321}}
\email{miguelcruz02@uv.mx}
\affiliation{Facultad de Física, Universidad Veracruzana 91097, Xalapa, Veracruz, México}

\author{Norman Cruz\orcidicon{0000-0002-0737-3497}}
\email{norman.cruz@usach.cl}
\affiliation{Departamento de Física, Universidad de Santiago de Chile, Avenida Victor Jara 3493, Estación Central, 9170124 Santiago, Chile}
\affiliation{Center for Interdisciplinary Research in Astrophysics and Space Exploration (CIRAS), Universidad de Santiago de Chile, Avenida Libertador Bernardo O’Higgins 3363, Estación Central, 9170022 Santiago, Chile}

\author{Esteban González\orcidicon{0000-0001-6769-5722}}
\email{esteban.gonzalez@ucn.cl}
\affiliation{Departamento de Física, Universidad Católica del Norte, Avenida Angamos 0610, Casilla 1280, Antofagasta, Chile}

\author{Samuel Lepe\orcidicon{0000-0002-3464-8337}}
\email{samuel.lepe@pucv.cl}
\affiliation{Instituto de Física, Facultad de Ciencias, Pontificia Universidad Católica de Valparaíso, Avenida Brasil 2950, Valparaíso, Chile.}

\begin{abstract}
In this work, we test the capability of an exact solution found in the framework of a nonlinear extension of the Israel-Stewart theory to fit the supernovae Ia, gravitational lensing, and black hole shadow data. This exact solution is a generalization of one previously found for a dissipative unified dark matter model in the context of the near-equilibrium description of dissipative processes, where we do not have the full regime of the nonlinear picture. This generalized solution is restricted to the case where a positive entropy production is guaranteed and is tested under the condition that ensures its causality, local existence, and uniqueness. From the observational constraints, we found that this generalized solution is a good candidate in the description of the observational late-time data used in this work, with best-fit values $H_{0}=73.2_{-0.9}^{+0.8}\,\frac{km/s}{Mpc}$, $q_{0}=-0.41_{-0.03}^{+0.03}$, $\hat{\xi}_{0}=0.88_{-0.17}^{+0.09}$, $\epsilon=0.34_{-0.04}^{+0.03}$, and $k=0.27_{-0.20}^{+0.37}$. Therefore, we show that the nonlinear regime of the Israel-Stewart theory consistently describes the recent accelerated expansion of the universe without the inclusion of some kind of dark energy component and also provides a more realistic description of the fluids that make up the late Universe.
\end{abstract}
\maketitle

\section{\label{sec:introduction} Introduction}
The current picture of the universe is inconceivable without a dark component, which leads to an accelerated cosmic expansion in the framework of general relativity, and it is usually named dark energy (DE). A first attempt to explain this phenomenon motivated the introduction of the cosmological constant in Einstein's field equations. This simple modification is well known as $\Lambda$CDM or standard cosmological model. However, despite its great success in fitting many of the sets of observational data, doesn't provide a compelling answer about the actual nature of the DE and it lacks physical consistency to describe some observational facts: the uniformity of the temperature for the photons coming from the cosmic microwave background (CMB) and the flattening behavior of the universe\footnote{In fact, some tensions on the actual value of the curvature parameter are well known in modern cosmology, see for instance Refs. \cite{DiValentino:2019qzk,Handley:2019tkm}.}, which are the horizon and flatness problems. A viable explanation for these issues was granted by the inflationary theory \cite{Guth:1980zm}, a period of over-accelerated expansion after the Big Bang. Another relevant problem is the value of the cosmological constant $\Lambda$. In the presence of a cosmological constant, the empty space has an energy density of the form, $\bar{\rho}_{emp} \rightarrow\rho_{emp}+\Lambda$. Observationally, this vacuum energy density is constrained to be smaller than $10^{-47}$ GeV, and from elementary theories of particles was found that such value must be around $10^{71}$ GeV \cite{Weinberg:1988cp}, representing a discrepancy of 118 orders of magnitude between theory and observations. Besides, from some other cosmological results, a very tiny (non-zero) and extremely fine-tuned value for $\bar{\rho}_{emp}$ can be found, $\bar{\rho}_{emp}\simeq\mathcal{O}(10^{-123})$ \cite{Padmanabhan:2002ji}.

On the other hand, assuming that observations are correct, the small value obtained for the density $\bar{\rho}_{emp}$ leads to confusing results. From the definition of the vacuum density given above, we can see that such density must remain constant throughout the cosmic evolution and, according to the $\Lambda$CDM description, the energy density associated with the matter sector typically has a decreasing behavior, $\rho_{m}\propto a^{-3}$, where $a$ is the scale factor. Therefore one could expect a difference of several orders of magnitude between both densities along cosmic expansion. However, the quotient $\rho_{m}/\bar{\rho}_{emp}$ is almost equal to one at present time, i.e., both components aggregate similar amounts to the total energy budget of the universe. This coincidence is not well understood and it is known as cosmological coincidence problem \cite{Velten:2014nra}. Another well-known problem is the $H_{0}$ tension, which presents a discrepancy of $5\sigma$ between the current value of the Hubble parameter inferred from Planck CMB (assuming the $\Lambda$CDM cosmology) and the obtained from local measurements of Cepheid (model-independent) \cite{Riess:2021jrx}. This tension is also supported by the $H_{0}$ Lenses in COSMOGRAIL's Wellspring (H0LiCOW) collaboration, with a discrepancy of $5.3\sigma$ concerning the value inferred from Planck CMB \cite{Wong:2019kwg}. These and some other problems indicated that new alternatives to describe the dark component of the universe are necessary. One possibility is given by dynamical DE models, as stated in Ref. \cite{Jaime:2018ftn}, and an interesting proposal can be found in \cite{Jaber:2021hho}. Within the Einstein framework, this means that this component can be thought of as a generic dynamical fluid with a given equation of state (EoS) that relates the pressure ($p$) with the energy density of the fluid. Recent results show that for a barotropic EoS $p=\omega\rho$, where $\omega$ is known as barotropic index, this component could have a $\omega$ parameter situated between the quintessence ($-1<\omega<-1/3$) and phantom ($\omega<-1$) regimes \cite{Planck:2018vyg}.

Going beyond the standard cosmological model to describe DE can be a difficult task, but an interesting approach to this problem is given by the inclusion of causal dissipative effects (bulk viscosity) in the fluid description \cite{Israel:1976tn,Israel:1979wp,Pavon:1990ha,Chimento:1993zc}. Dissipative processes can characterize important stages of cosmic evolution: reheating of the universe, decoupling of neutrinos from the cosmic plasma, and nucleosynthesis. Besides, viscosity can also be present in several astrophysical mechanisms as, for example, the collapse of radiating stars to a neutron star or black hole and in the accretion of matter around neutron stars or black holes \cite{Maartens:1996vi}. A possible origin for bulk viscosity is attributed to the existence of mixtures. In the single fluid description, the universe as a whole can be characterized by the particle number density, $n=n_{1}+...+n_{i}$, therefore the simple assumption of different cooling rates in the expanding mixture can lead to a non-vanishing viscous pressure \cite{Zimdahl:1996fj}. As an effect of bulk viscosity, the kinetic energy of particles is converted into heat, thus a reduction of the effective pressure of the fluid is expected, and the condition for the Hubble parameter $H\equiv\dot{a}/a>0$ is supported by $\Pi\leq 0$ \cite{Maartens:1996vi}, being $\Pi$ the bulk viscous pressure. An interesting alternative to explain the origin of bulk viscosity was explored in \cite{Wilson:2006gf,Mathews:2008hk}, where was shown that the decay of DM into relativistic particles allow naturally the emergence of dissipative effects in the cosmic fluid. A scheme such as the aforementioned could help to understand more clearly the role of bulk viscosity in the cosmic expansion. Recent results show that a decaying scenario for DM increases the expansion rate relative to $\Lambda$CDM and such behavior provides an alleviation for the $H_{0}$ and $\sigma_{8}$ tensions \cite{Pandey:2019plg}. In the standard description, a cosmological fluid has a constant temperature and can not generate entropy or {\it frictional} heating. Therefore, the cosmological constant as a source of DE has no sense at a thermodynamical level \cite{Cardenas:2018nem}. Also, within the standard description, DE is not allowed to interact with other components of the universe. However, a more realistic (and consistent) picture of the universe should not forbid such interaction \cite{Wang:2016lxa}. Several works demonstrate that under the description of thermodynamics for reversible processes, the dissipative effects play a relevant role at late-times of cosmic evolution and, even more, are capable of conceding a phantom cosmology under some assumptions \cite{Cataldo:2005qh,Disconzi:2014oda,Cruz:2017lbu,Cruz:2016rqi,Cardenas:2020exv}. Beyond the bounds of reversible processes, the dissipative effects can be used to solve some problems at the thermodynamics level of the phantom regime. For instance, the simultaneous positivity of entropy and temperature \cite{Cruz:2018arw}.

An interesting work on viscous cosmology can be found in Ref. \cite{Bemfica:2019cop}, where the authors discuss that when dissipative effects are coupled to the gravitational sector the causal structure of the theory can be maintained without invoking the near equilibrium condition or a specific EoS, besides is the first work that establishes the initial-value problem in the context of viscous fluids. This formulation represents an important ingredient to perform numerical simulations. This last result could provide a way to assert or discard the findings of Ref. \cite{Alford:2017rxf}, where was claimed that bulk viscosity could contribute significantly to the emission of gravitational waves in neutron star mergers. Some other results can be found in Refs. \cite{Barta:2019tpv,BravoMedina:2019han}, where the role of bulk viscosity is studied in other contexts such as the radial oscillation of relativistic stars and the cosmological implications for universes filled with Quark-Gluon plasma. Recent studies show that bulk viscous cosmologies are not ruled out by the observational data at all. In fact, in Ref. \cite{Yang:2019qza} the bulk viscosity effects were tested with the combination of data coming from CMB, type Ia supernovae (SNe Ia), and observations of the Hubble parameter (OHD). This complete study shows that the parameter that characterizes the viscosity must be non-zero and, in addition, the results obtained seem to relieve the current $H_0$ tension. Somehow this seems to be consistent with the results of Ref. \cite{Poulin:2018cxd}, where was proposed that the $H_{0}$ tension might be resolved in the context of a new theory and consists in the annexation of an extra form of DE. In the interesting compendium \cite{DiValentino:2021izs}, bulk viscous effects are explored as a viable alternative to relieve the $H_{0}$ tension (see also Ref. \cite{Normann:2021bjy}).

Our aim in this paper is to study the capability to fit the observational SNe Ia, gravitational lensing, and black hole shadows (BHS) data by an exact solution found in the nonlinear regime of the Israel-Stewart theory in a flat Friedmann–Lemaître–Robertson–Walker (FLRW) metric. Such solution represents a generalization of one previously found in the context of the linear regime for a unified model of dissipative dark matter (DM) \cite{Cruz:2018psw}. Since no other fluid was included to obtain this solution, it is suitable only to describe the late-time evolution. Furthermore, according to \cite{Cruz:2019uya}, the accelerated expansion in the linear solution can be described within a range of the parameters of the model for a pressureless DM fluid, which is compatible with OHD and SNe Ia data. The main motivation to consider the nonlinear extension of the Israel-Stewart theory comes from the fact that the aforementioned exact solution represents an accelerated universe, which means that the near equilibrium condition is violated, i.e., the viscous stress denoted by $\Pi$ is greater than the equilibrium pressure $p$ of the dissipative DM fluid. This violation holds whenever the accelerated expansion is due only to the negative pressure of the viscous fluid \cite{Maartens:1995wt}. Nevertheless, by relaxing the near equilibrium condition it was found that a nonlinear extension of the Israel-Stewart model ensures a bounded value for $\left|\Pi\right|/p$ as well as the fulfillment of the second law under the upper bound on $\Pi$ \cite{Maartens:1996dk}. Therefore,  this approach is more adequate to describe an accelerated cosmic evolution without the inclusion of some DE component. In this description, high values in the nonadiabatic contribution to the speed of sound can be allowed. On the other hand, in this work we focus on the positive entropy production case within the nonlinear extension. This thermodynamic restriction leads to a simplified version of the full theory. Therefore, the consideration of the full theory could reveal some other different aspects from those obtained here.

This work is organized as follows: In Section \ref{sec:generalities}, we provide a brief description of the Israel-Stewart model and discuss the passage from this scheme to the nonlinear regime of dissipative cosmology in Section \ref{sec:solution}. The exact solutions emerging from both scenarios are discussed. Section \ref{sec:fitting} is devoted to briefly explain the procedure for the observational constraints, whereas in Sections \ref{sec:SNeIa}, \ref{sec:lensing}, and \ref{sec:BHS}, we describe the construction of the merit function for the SNe Ia, gravitational lensing, and BHS data, respectively; while in Section \ref{sec:Priors} we present the prior used in the constraint and the implications of the causality, local existence and uniqueness condition of the Israel-Stewart theory in the full nonlinear regime. In Section \ref{sec:results}, we present the results obtained for the observational analysis of the nonlinear solution. Finally, in Section \ref{sec:conclusions}, we give some final comments and remarks. In this work, we will consider $8\pi G=c=1$ units, except in the Section \ref{sec:fitting}.

\section{\label{sec:generalities} Preliminaries of the Israel-Stewart theory}
In general relativity, within the context of the Israel-Stewart's framework for a flat FLRW metric without a cosmological constant, the energy density of the viscous fluid obeys the continuity equation $\dot{\rho}+3H[(1+\omega)\rho+\Pi]=0$, were we have considered a barotropic EoS of the form $p =\omega\rho$. On the other hand, the energy density of the viscous fluid can be written in terms of the Hubble parameter through the Friedmann equation $3H^{2}=\rho$. Also note that, if we consider the acceleration equation $\dot{H}+H^{2}=-[\rho +3(p+\Pi)]/6$, then we can obtain the explicit form of the bulk viscous pressure as
\begin{equation}
\Pi=-2\dot{H}-3H^{2}(1+\omega).
\label{viscouspressure}
\end{equation}
Besides, in the full Israel-Stewart theory, the bulk viscous pressure obeys the equation \cite{Israel:1979wp}
\begin{equation}
\tau \dot{\Pi}+\left(1+\frac{1}{2}\tau\Delta\right)\Pi=-3\xi(\rho)H,
\label{eq:pi}	
\end{equation}
where in the limit $\tau\rightarrow 0$, we have the non-causal Eckart's theory \cite{Eckart:1940te}. This latter case is not considered in this work and we will define $\Delta$ later in this section. Therefore, the Hubble parameter for a universe dominated only by a dissipative fluid obeys the following transport equation \cite{Cruz:2016rqi,Cruz:2017bcv}
\begin{widetext}
\begin{equation}
\ddot{H}+\left[3H(1+\omega)+\frac{\Delta}{2} \right]\dot{H}+\frac{9}{2}\epsilon(1-\omega^{2})\left[\frac{(1+\omega)}{3^{s}\xi_{0}}H^{1-2s}-1\right]H^{3}+\frac{\epsilon (1-\omega^{2})}{3^{s-1}\xi_{0}}\dot{H}H^{2(1-s)}+\frac{3}{4}(1+\omega)\Delta H^{2}=0,
\label{eq:trans}
\end{equation}
\end{widetext}
where the dot denotes derivatives with respect to the cosmic time. Some remarks are in order: $\epsilon$ is a constant parameter that accounts for the causality of the model, the bulk viscosity has been chosen proportional to the energy density of the dissipative fluid of the form $\xi=\xi_{0}\rho^{s}$, being $\xi_{0}$ a positive constant and $s$ an arbitrary real parameter. An expanding evolution can take place in a scenario for a universe with viscous DM and a cosmological constant, see for instance Ref. \cite{Cruz:2018yrr}; de Sitter solutions are allowed by such model for some specific values of the parameter $s$, being $s=1/2$ one of them. Although the above election for the bulk viscosity is widely used in the literature, it is still arbitrary. The election $s=1/2$ also has the important property of simplifying Eq. (\ref{eq:trans}) to integrate it, in which we defined
\begin{equation}
\Delta := 3H + \frac{\dot{\tau}}{\tau} - \frac{\dot{T}}{T}-\frac{\dot{\xi}}{\xi},
\label{eq:delta}	
\end{equation}
being $T$ the barotropic temperature, which is generally written as $T=T_{0}(\rho/\rho_{0})^{\omega/(1+\omega)}$, by means of the Gibbs integrability condition and $T_{0}$ is the value of the temperature when $\rho = \rho_{0}$. Therefore, the elected value for the parameter $s$ represents interesting physical scenarios; see for instance Refs. \cite{Cruz:2016rqi,Cruz:2017bcv,Cruz:2017lbu}, where some aspects of the observed universe such as phantom regime and/or accelerated cosmic expansion were studied in the framework of viscous cosmology.

It is important to mention that, from the causality condition, we have the restriction $0\leq\omega<1$, which can be obtained from the definition of the linear relaxation time \cite{Maartens:1996vi}
\begin{equation}
\tau = \frac{\xi}{c^{2}_{b}(\rho + p)} = \frac{\xi_{0}}{\epsilon(1-\omega^{2})}\rho^{s-1},
\label{eq:tau}
\end{equation}   
where $c^{2}_{b}$ is the non adiabatic contribution to the speed of sound $v$ (or speed of bulk viscous perturbations). In this sense, the causality condition reads $v^{2} = c^{2}_{s}+c^{2}_{b} \leq 1$, being $c^{2}_{s}:=(\partial p/\partial\rho)$ the adiabatic contribution. Therefore, to fulfill the causality condition, we must have $c^{2}_{b}=\epsilon (1-\omega)$, where the parameter $\epsilon$ was introduced and must satisfy the condition $0<\epsilon\leq 1$. The previous equation for the relaxation time corrects the relation $\xi/\rho\tau=1$, which is considered for simplicity in several works.

Taking $s=1/2$ in Eq. (\ref{eq:trans}), this expression becomes
\begin{equation}
\ddot{H}+ b_{1}H\dot{H}- \Gamma H^{-1}\dot{H}^{2}+ b_{2} H^{3} = 0,
\label{eq:good}
\end{equation}
where, for simplicity in the notation, we have defined the constant coefficients
\begin{subequations}
\begin{equation}
b_{1} =3\left\lbrace 1+\frac{\epsilon \left(1-\omega^{2}\right)}{\sqrt{3}\xi_{0}}\right\rbrace,
\label{eq:constant1}
\end{equation}
\begin{equation}
b_{2} =\frac{9}{4}(1+\omega)\left\lbrace 1-2\epsilon(1-\omega)+\frac{2\epsilon(1-\omega^{2})}{\sqrt{3}\xi_{0}}\right\rbrace,
\label{eq:constant2}
\end{equation}
\begin{equation}
\Gamma =\frac{(1+2\omega)}{(1+\omega)}.
\label{eq:constant3}
\end{equation}
\end{subequations}
It is worth mentioning that if we consider the limit case $\omega = 0$ in the differential equation written in (\ref{eq:good}), we obtain the non-relativistic or DM case ($\gamma=1$) explored in Ref. \cite{Mohan:2017poq}.

\subsection{\label{sec:solution} Exact solution and its extension to a nonlinear regime}
In the following section, we will provide some generalities of the exact solution found in Ref. \cite{Cruz:2018psw}. We refer the reader to the aforementioned reference to see the technical details. In this solution, the DM is the only fluid component in the universe's energy density budget. It also experiences a dissipative process in the form of bulk viscosity during its cosmic evolution. Then, by introducing the change of variable $x = \ln (a/a_{0})$ in Eq. (\ref{eq:good}), where $a_{0}$ is the value for the scale factor at which the Hubble parameter becomes $H_{0}$, one gets
\begin{equation}
\frac{d^{2}H}{dx^{2}}+b_{1}\frac{dH}{dx}+\frac{1-\Gamma}{H}\left(\frac{dH}{dx}\right)^{2}+b_{2}H = 0.
\label{eq:Hx}
\end{equation}
Therefore, solving the previous equation, considering that the scale factor are related with the redshift $z$ through the expression $1+z=a_{0}/a$, we can obtain the Hubble parameter as a function of $z$, as follows
\begin{equation}
H(z) = H_{0}\frac{(1+z)^{\alpha}\cosh^{(1+\omega)}{\left[\beta(\ln{(1+z)}+C)\right]}}{\cosh^{(1+\omega)}{(\beta C)}},
\label{eq:newHubble}
\end{equation}
where
\begin{subequations}
\begin{equation}
C = \frac{1}{\beta}\arctanh{\left[\frac{(q_{0}+1)-\alpha}{(1+\omega)\beta}\right]},
\end{equation}
\begin{equation}
\alpha = \frac{\sqrt{3}(1+\omega)}{2\xi_{0}}\left[\sqrt{3}\xi_{0}+\epsilon(1+\omega)(2-(1+\omega))\right] ,
\end{equation}
\begin{equation}
\beta = \frac{\sqrt{3}}{2\xi_{0}}\sqrt{6\xi_{0}^{2}\epsilon(2-(1+\omega))+\epsilon^{2}(1+\omega)^{2}(2-(1+\omega))^{2}}.
\end{equation}
\end{subequations}
The form of the constants defined previously come from the initial conditions for the Hubble parameter and its derivative, $H(z=0)=H_{0}$ and $H'(z=0)=q_{0}$, respectively; where $z=0$ represents the current time, prime denotes derivative with respect to $z$, and $q_{0}$ is the deceleration parameter at the current time\footnote{The redshift (or scale factor) derivative of the Hubble parameter can be related to the deceleration parameter through the relation $\frac{dH(z)}{dz}=\frac{1+q(z)}{1+z}H(z)$.}. In this description, the value $\omega=0$ (CDM) can be considered as the limit case to perform a comparison with the $\Lambda$CDM model. In Ref. \cite{Cruz:2019uya} the solution (\ref{eq:newHubble}) was tested with the use of the joint SNe Ia+OHD data. 

The Israel-Stewart model is based on the assumption that only small deviations from equilibrium are permitted. This implies that we must have a rapid adjustment to the cooling caused by the cosmic expansion, i.e., $\tau H \ll 1$, which can be seen from Eq. (\ref{eq:tau}).  However, by using SNe Ia data some drawbacks in the model were found, according to the analysis performed in \cite{Cruz:2019uya}. In order to have late-times cosmic expansion, the best-fit parameters indicate that $\xi_{0} \gg 1$ together with $\omega \approx 0$. Therefore, the rapid adjustment condition is lost. Nevertheless, to explore dissipative processes beyond the near equilibrium condition mentioned before, it was introduced a nonlinear extension of the Israel-Stewart model in Ref. \cite{Maartens:1996dk}, in which large deviations from equilibrium are allowed. In this case, the transport equation for the viscous pressure given in (\ref{eq:pi}) transforms into
\begin{equation}
\begin{split}
& 3\xi(\rho)H+\tau \dot{\Pi}\left(1+\frac{\tau_{*}}{\xi(\rho)}\Pi \right) + \\
& \left[1+\frac{1}{2}\tau \Delta \left(1+\frac{\tau_{*}}{\xi(\rho)}\Pi \right)+3\tau_{*}H \right]\Pi = 0,
\label{eq:pi2}	
\end{split}
\end{equation}
where $\tau_{*} \geq 0$ characterizes the time scale for the nonlinear effects. The must simple assumption for this time scale is given by $\tau_{*}=k^{2}\tau$. Then, for $k=0$, we recover the linear Israel-Stewart theory, ensuring the causality of the model. 

Using the transport equation (\ref{eq:pi2}), together with a FLRW metric, and repeating the procedure as done previously for the linear Israel-Stewart theory, we arrive at the following, more complicated second-order differential equation for the Hubble parameter \cite{Cruz:2017lbu,Maartens:1996dk}
\begin{widetext}
\begin{equation}
\begin{split}
& \left[1-\frac{k^{2}}{\epsilon(1-\omega)}-\left(\frac{2k^{2}}{3\epsilon(1-\omega^{2})} \right)\frac{\dot{H}}{H^{2}} \right]\left\lbrace \ddot{H}+3H\dot{H}+\left(\frac{1-2(1+\omega)}{(1+\omega)} \right)\frac{\dot{H}^{2}}{H}+\frac{9}{4}(1+\omega) H^{3} \right\rbrace \\
& + \frac{3\epsilon(1-\omega^{2})}{2\sqrt{3}\xi_{0}}\left[1+\left(\frac{\sqrt{3}\xi_{0} k^{2}}{\epsilon(1-\omega^{2})}\right)H^{2s-1} \right]H^{2(1-s)}(2\dot{H}+3(1+\omega) H ^{2})-\frac{9}{2}\epsilon(1-\omega^{2})H^{3} = 0,
\label{eq:brigida}
\end{split}
\end{equation}
\end{widetext}
which describes the behavior of the model in the nonlinear regime of Israel-Stewart's theory for an arbitrary $s$, in an analogous way as Eq. \eqref{eq:trans} does for the linear regime. On the other hand, by considering the special case $s=1/2$, in Ref. \cite{Chimento:1997ga} was found that for models with barotropic temperature, the term $\dot{H}/H^{2}$ appearing in the differential equation (\ref{eq:brigida}) must be bounded together with $k^{2}<1$ in order to have a positive entropy production. Under these assumptions the resulting second order differential equation for $H$ at first order in $k^{2}$ takes the same structure as Eq. (\ref{eq:good}) but, in this case, the constant coefficients $b_{1}$, $b_{2}$ and $\Gamma$ are redefined as follows   
\begin{subequations}
\begin{equation}
\bar{b}_{1} = b_{1} + 2\sqrt{3}\frac{k^{2}(1+\omega)}{\xi_{0}},
\end{equation}
\begin{equation}
\bar{b}_{2} = b_{2} + \frac{3\sqrt{3}}{2}\frac{k^{2}(1+\omega)^{2}}{\xi_{0}},
\end{equation}
\begin{equation}
\bar{\Gamma} = \Gamma -\frac{2k^{2}}{\sqrt{3}\xi_{0}}.
\end{equation}
\end{subequations}
Therefore, the solution for the Hubble parameter in this nonlinear regime will have the same mathematical structure as the solution given in (\ref{eq:newHubble}) for the linear regime, but will describe dissipative processes far from equilibrium. Also note that within the nonlinear regime, the case $\omega=0$ leads to $\bar{\Gamma} = 1 -\frac{2k^{2}}{\sqrt{3}\xi_{0}}$, which maintains the structure of the differential equation (\ref{eq:Hx}). Then, this limit case can be considered in the context of the solution discussed above. Therefore, in terms of the redshift, the Hubble parameter for the nonlinear regime can be penned as follows
\begin{equation}
H(z) = \frac{H_{0}(1+z)^{\kappa\bar{\alpha}}\cosh^{\kappa(1+\omega)}{\left[\bar{\beta}(\ln{(1+z)}+\bar{C})\right]}}{\cosh^{\kappa(1+\omega)}{(\bar{\beta}\bar{C}})}, 
\label{eq:nonlinearHubble}
\end{equation}
with
\begin{subequations}
\begin{equation}\label{defofbC}
\bar{C} = \frac{1}{\bar{\beta}}\arctanh{\left[\frac{(q_{0}+1)-\kappa\bar{\alpha}}{\kappa(1+\omega)\bar{\beta}}\right]},
\end{equation}
\begin{widetext}
\begin{equation}
\bar{\alpha} = \frac{\sqrt{3}(1+\omega)}{2\xi_{0}}\left[\sqrt{3}\xi_{0}+\epsilon(1+\omega)(2-(1+\omega))+2k^{2}(1+\omega)\right],
\end{equation}
\begin{equation}
\bar{\beta} = \frac{\sqrt{3}}{2\xi_{0}}\sqrt{6\xi_{0}^{2}\epsilon(2-(1+\omega))+\epsilon^{2}(1+\omega)^{2}(2-(1+\omega))^{2}-4k^{2}(1+\omega)\epsilon(2-(1+\omega))\left(\sqrt{3}\xi_{0}+2(1+\omega)\right)},
\end{equation}
\end{widetext}
\begin{equation}
\kappa = \frac{\sqrt{3}\xi_{0}}{\sqrt{3}\xi_{0}+2k^{2}(1+\omega)}.
\end{equation}
\end{subequations}
Note that in the previous expressions the value $k=0$ still represents the model in the linear regime of the Israel-Stewart's theory since, for such case, we have $\bar{\alpha}=\alpha$, $\bar{\beta}=\beta$, $\bar{C}=C$ and $\kappa=1$.

\section{\label{sec:fitting} Observational constraints}
In this section, we shall constrain the free parameters of the nonlinear solution with the SNe Ia, gravitational lensing, and BHS data. For this purpose, we compute the best-fit parameters of the solution at $1\sigma \,(68.3\%)$ of confidence level (CL) with the affine-invariant Markov chain Monte Carlo (MCMC) method~\cite{Goodman_Ensemble_2010}, implemented in the pure-Python code \textit{emcee} \cite{Foreman-Mackey:2012any}. Following this line, for this MCMC analysis, we need to construct the following Gaussian likelihood:
\begin{equation}\label{likelihood}
\mathcal{L}_{I}\propto\exp{\left(-\frac{\chi_{\text{I}}^{2}}{2}\right)},
\end{equation}
where $\chi_{\text{I}}^{2}$ is the merit function and $I$ stands for each data set, namely SNe Ia, gravitational lensing, BHS, and their joint analysis where $\chi^{2}_{\text{joint}}=\sum{\chi^{2}_{I}}$. Therefore, in the following subsections, we will briefly describe the construction of the merit function of each data set.

\subsection{\label{sec:SNeIa} Type Ia supernovae}
For the SNe Ia data, we consider the Pantheon+ sample \citep{Brout:2022vxf}, which consists in 1701 data points in the redshift range $0.001\leq z\leq 2.26$, whose respective merit function can be conveniently constructed in matrix notation (denoted by bold symbols) as
\begin{equation}\label{meritSNe}
\chi_{\text{SNe}}^{2}=\mathbf{\Delta D}(z,\boldsymbol{\theta},M)^{\dagger}\mathbf{C}^{-1}\mathbf{\Delta D}(z,\boldsymbol{\theta},M),
\end{equation}
where $\left[\mathbf{\Delta D}(z,\boldsymbol{\theta},M)\right]_{i}= m_{B,i}-M-\mu_{th}(z_{i},\boldsymbol{\theta})$ and $\mathbf{C}=\mathbf{C}_{\text{stat}}+\mathbf{C}_{\text{sys}}$, with $\mathbf{C}$ the total uncertainty covariance matrix. The matrices $\mathbf{C}_{\text{stat}}$ and $\mathbf{C}_{\text{sys}}$ account for the statistical and systematic uncertainties, respectively, and $\boldsymbol{\theta}$ encompasses the free parameters of the model. The quantity $\mu_{i}=m_{B, i}-M$ corresponds to the observational distance modulus of the Pantheon+ sample, which is obtained by a modified version of Trip's formula \citep{Tripp:1997wt} and the BBC (BEAMS with Bias Corrections) approach \citep{Kessler:2016uwi}; while $m_{B, i}$ is the corrected apparent B-band magnitude of a fiducial SNe Ia at redshift $z_{i}$, and $M$ is the fiducial magnitude of a SNe Ia, which must be jointly estimated with the free parameters of the model under study.

The theoretical  distance modulus for a spatially flat FLRW spacetime is given by
\begin{equation}\label{theoreticaldistance}
\mu_{th}(z_{i},\boldsymbol{\theta})=5\log_{10}{\left[\frac{d_{L}(z_{i},\boldsymbol{\theta})}{\text{Mpc}}\right]}+25,
\end{equation}
with $d_{L}(z_{i},\boldsymbol{\theta})$ the  luminosity distance given by
\begin{equation}\label{luminosity}
d_{L}(z_{i},\boldsymbol{\theta})=c(1+z_{i})\int_{0}^{z_{i}}{\frac{dz'}{H_{th}(z',\boldsymbol{\theta})}},
\end{equation}
where $c$ is the speed of light given in units of $\text{km/s}$.

In principle, there is a degeneration between $M$ and $H_{0}$. Hence, to constraint $H_{0}$ using SNe Ia data alone, it is necessary to include the SH0ES (Supernovae and $H_{0}$ for the Equation of State of the dark energy program) Cepheid host distance anchors, with a merit function of the form
\begin{equation}\label{Cepheidmerit}
\chi^{2}_{\text{Cepheid}}=\mathbf{\Delta D}_{\text{Cepheid}}\left(M\right)^{\dagger}\textbf{C}^{-1}\mathbf{\Delta D}_{\text{Cepheid}}\left(M\right),
\end{equation}
where 
$\left[\mathbf{\Delta D}_{\text{Cepheid}}\left(M\right)\right]_{i}=\mu_{i}\left(M\right)-\mu_{i}^{\text{Cepheid}}$,  with $\mu_{i}^{\text{Cepheid}}$ the Cepheid calibrated host-galaxy distance obtained by SH0ES \citep{Riess:2021jrx}. So, we use the Cepheid distances as the ``theory model'' to calibrate $M$, considering that the difference $\mu_{i}\left(M\right)-\mu_{i}^{\text{Cepheid}}$ is sensitive to $M$ and largely insensitive to other parameters of the cosmological model. Considering the total uncertainty covariance matrix for Cepheid is contained in the total uncertainty covariance matrix $\mathbf{C}$, we can define the merit function for the SNe Ia data as
\begin{equation}\label{SNemeritfull}
\chi_{\text{SNe}}^{2}=\mathbf{\Delta D'}(z,\boldsymbol{\theta},M)^{\dagger}\mathbf{C}^{-1}\mathbf{\Delta D'}(z,\boldsymbol{\theta},M),
\end{equation}
where
\begin{equation}\label{SNeresidual}
\Delta\mathbf{D'}_{i}=\left\{\begin{array}{ll}
m_{B,i}-M-\mu_{i}^{\text{Cepheid}} & i\in\text{Cepheid host} \\
\\ m_{B,i}-M-\mu_{th}(z_{i},\boldsymbol{\theta}) & \text{otherwise}
\end{array}
\right..
\end{equation}

From now on, we will omit the nuisance parameter $M$, for which we consider the flat prior $-20<M<-18$ in our MCMC analysis, and we will focus our analysis only on the free parameters of the cosmological model.

\subsection{\label{sec:lensing} Gravitational lensing}
When a background object (the source) is lensed due to the gravitational force of an intervening massive body (the lens), it is obtained as a result the generation of multiple images. Therefore, the light rays emitted from the source will take different paths through the space-time at the different image positions arriving at the observer at different times. In this sense, the time delay of two different images $k$ and $l$ depends on the mass distribution along the line of sight of the lensing object and can be calculated as
\begin{equation}\label{lensing}
    \Delta t_{kl}=\frac{D_{\Delta t}}{c}\left[\frac{\left(\phi_{k}-\beta\right)^{2}}{2}-\psi(\phi_{k})-\frac{\left(\phi_{l}-\beta\right)^{2}}{2}+\psi(\phi_{l})\right],
\end{equation}
where $\phi_{k}$ and $\phi_{l}$ are the angular position of the images, $\beta$ is the angular position of the source, $\psi(\phi_{k})$ and $\psi(\phi_{l})$ are the lens potential at the image positions, and $D_{\Delta t}$ is the ``time-delay distance'', which is theoretically given by the expression \cite{Treu:2016ljm}
\begin{equation}\label{time-delay}
    D_{\Delta t}^{th}(\mathbf{z},\boldsymbol{\theta})=\left(1+z_{l}\right)\frac{d_{A,l}(z_{l},\boldsymbol{\theta})d_{A,s}(z_{s},\boldsymbol{\theta})}{d_{A,ls}(z_{ls},\boldsymbol{\theta})},
\end{equation}
where $l$, $s$, and $ls$ stand for the lens, the source, and between the lens and the source, respectively, $\mathbf{z}=(z_{l},z_{s},z_{ls})$, and $d_{A,j}$ is the angular diameter distance, which can be written in terms of the luminosity distance \eqref{luminosity} as $d_{L}(z_{j},\boldsymbol{\theta})=d_{A,j}(1+z_{j})^{2}$ or
\begin{equation}\label{angulardistance}
    d_{A,j}(z_{j},\boldsymbol{\theta})=\frac{c}{(1+z_{j})}\int_{0}^{z_{j}}{\frac{dz'}{H_{th}(z',\boldsymbol{\theta})}}.
\end{equation}

In this paper, we consider the gravitational lensing compilation provided by the H0LiCOW collaboration \cite{Wong:2019kwg}, which consist in six lensed quasars: B1608+656 \cite{Jee:2019hah}, SDSS 1206+4332 \cite{Birrer:2018vtm}, WFI2033-4723 \cite{Rusu:2019xrq}, RXJ1131-1231 , HE 0435-1223, and PG 1115-080 \cite{Chen:2019ejq}; whose respective merit function can be constructed as
\begin{equation}\label{H0LiCOWmerit}
    \chi^{2}_{\text{H0LiCOW}}=\sum_{i=1}^{6}{\left[\frac{D_{\Delta t,i}-D_{\Delta t}^{th}(\mathbf{z}_{i},\boldsymbol{\theta})}{\sigma_{D_{\Delta t},i}}\right]^{2}},
\end{equation}
where $D_{\Delta t,i}$ is the observational time-delay distance of the lensed quasar at redshift $\mathbf{z}_{i}=(z_{l,i};z_{s,i};z_{ls,i})$ with an associated error $\sigma_{D_{\Delta t},i}$ (for more details see Ref. \cite{Wong:2019kwg}). It is important to note that, for $z\to 0$ the angular diameter distance \eqref{angulardistance} tends to $d_{A}\to cz/H_{0}$ and, therefore, the gravitational lensing data of the H0LiCOW collaboration is sensitive to $H_{0}$, with a weak dependency on other cosmological parameters.

\subsection{\label{sec:BHS} Black hole shadows}
The BHS data is of interest to study our local universe since their dynamic is quite simple and can be seen as standard rulers if the angular size redshift $\alpha$, the relation between the size of the shadow and the mass of the supermassive black hole that produces it, is established \cite{Escamilla-Rivera:2022mkc}. The first detection of a BHS was made by the Event Horizon Telescope Collaboration for the M87* supermassive black hole \cite{EventHorizonTelescope:2019dse}, which was later complemented with the detection of Sagittarius A* (Sgr A*) \cite{EventHorizonTelescope:2022wkp}.

In a black hole (BH), light rays curve around its event horizon, creating a ring with a black spot center, the so-called shadow of the BH. The angular radius of the BHS for a Schwarzschild (SH) BH at redshift $z_{i}$ is given by
\begin{equation}\label{angularradius}
    \alpha_{SH}\left(z_{i},\boldsymbol{\theta}\right)=\frac{3\sqrt{3}m}{d_{A}(z_{i},\boldsymbol{\theta})},
\end{equation}
where $d_{A}(z_{i},\boldsymbol{\theta})$ is given by Eq. \eqref{angulardistance} (note that the sub-index $j$ is not necessary in this case) and $m=GM_{BH}/c^{2}$ is the mass parameter of the BH, with $M_{BH}$ the mass of the BH in solar masses units and $G$ the gravitational constant. 

It is common to write Eq. \eqref{angularradius} in terms of the shadow radius $\alpha_{SH}(z_{i},\boldsymbol{\theta})=R_{SH}/d_{A}(z_{i},\boldsymbol{\theta})$, where $R_{SH}=3\sqrt{3}GM_{BH}/c^{2}$ (the speed of light is given in units of $\text{m/s}$ in this case). Therefore, the merit function for the BHS data can be constructed as
\begin{equation}\label{BHSmerit}
    \chi^{2}_{BHS}=\sum_{i=1}^{2}{\left[\frac{\alpha_{i}-\alpha_{SH}(z_{i},\boldsymbol{\theta})}{\sigma_{\alpha,i}}\right]^{2}},
\end{equation}
where $\alpha_{i}$ is the observational angular radius of the BHS at redshift $z_{i}$ with an associated error $\sigma_{\alpha,i}$. It is important to note that for $z\to 0$ the angular radius \eqref{angularradius} tends to $\alpha_{SH}\to R_{SH}H_{0}/cz$ and, therefore, as well as in the gravitational lensing data, the BHS data is sensitive to $H_{0}$, with a weak dependency on other cosmological parameters. On the other hand, we divide Eq. \eqref{angularradius} by a factor of $1.496\times 10^{11}$ to obtain $\alpha_{SH}$ in units of $\mu as$.

\subsection{\label{sec:Priors} Priors and causality, local existence, and uniqueness condition}
The model is contrasted with the cosmological data through the Hubble parameter of the nonlinear solution given in Eq. (\ref{eq:nonlinearHubble}), whose respective parameter space is $\boldsymbol{\theta}=\left(H_{0},q_{0},\xi_{0},\epsilon,k,\omega\right)$. Since we are interested in the viability of our model to describe the recent accelerated expansion of the universe without the inclusion of a cosmological constant as the $\Lambda$CDM model, we assume a pressureless fluid ($\omega=0$) for the DM and a flat prior on $q_{0}$ of the form $-1<q_{0}<0$. The Hubble parameter at the current time is written in terms of the reduced Hubble constant $H_{0}=100\frac{km/s}{Mpc}h$, within the flat prior $0.55<h<0.95$. Considering also that $k^{2}< 1$ and $\xi_{0}>0$, then we use a flat prior on $k$ of the form $0<k<1$ and we make the change of variable $\xi_{0}=\hat{\xi}_{0}/(1-\hat{\xi}_{0})$ in order to simplify the sample of the full parameter space, with a flat prior on $\hat{\xi}_{0}$ of the form $0<\hat{\xi}_{0}<1$.

Following Ref. \cite{Bemfica:2019cop}, in order to ensure the causality of the Israel-Stewart theory in the fully nonlinear regime, we need to fulfill the condition
\begin{equation}\label{nonlinearcausality}
    \left[\frac{\xi}{\tau}+n\left(\frac{\partial p}{\partial n}\right)_{\rho}\right]\frac{1}{\rho+p+\Pi}\leq 1-\left(\frac{\partial p}{\partial\rho}\right)_{n},
\end{equation}
where $n$ is the baryon density. In our case, the universe is dominated only by a dissipative CDM, so we have a negligible baryon density and the above expression becomes (for illustrative purposes, we will keep $p=\omega\rho\neq 0$)
\begin{equation}\label{causality}
    \frac{\xi}{\tau\left(\rho+p+\Pi\right)}\leq 1-c_{s}^{2},
\end{equation}
which is reduced to the standard condition for causality and stability in the linearized regime given by Eq. \eqref{eq:tau} under the near equilibrium condition $\left|\Pi/\left(\rho+p\right)\right|\ll 1$ \cite{Denicol:2008ha,Pu:2009fj,Romatschke:2009im}. On the other hand, to express the equations of the IS theory as a first-order symmetric hyperbolic system, which implies the local existence and uniqueness of its solutions \cite{Reula:1995wk}, we need to fulfill the condition
\begin{equation}\label{FOSHcondition}
    \frac{\xi}{\tau\left(\rho+p+\pi\right)}+\left(\frac{\partial p}{\partial\rho}\right)_{n}+n\left(\frac{\partial p}{\partial n}\right)_{\rho}\frac{1}{\left(\rho+p+\Pi\right)}\geq 0,
\end{equation}
which for a negligible baryon density becomes
\begin{equation}\label{FOSH}
    -c_{s}^{2}\leq\frac{\xi}{\tau\left(\rho+p+\pi\right)}.
\end{equation}
Therefore, for $\omega=0$ (which implies $c_{s}=0$), we can ensure the causality, local existence, and uniqueness of our solution in the full nonlinear regime of the Israel-Stewart theory by the fulfillment of the conditions \eqref{causality} and \eqref{FOSH}, which can be condensed in only one condition of the form
\begin{equation}\label{maincondition}
    0\leq\frac{\xi}{\tau\left(\rho+\Pi\right)}\leq 1.
\end{equation}

From Eq. \eqref{eq:tau}, we can see that the condition \eqref{maincondition} only impose a constraint when $\rho>-\Pi$ of the form $0<\epsilon\leq 1+\Pi/\rho$, which, after some algebra, can be written as
\begin{equation}\label{epsiloncons}
    0<\epsilon\leq\frac{2}{3}\left(1+q\right).
\end{equation}
Thus, the parameter $\epsilon$, which was related to the causality in the linearized regime, is now upper bounded by the deceleration parameter $q$, which leads to some physical implications: 1) it is not possible that $q$ can take the value of $-1$ because this implies a solution without dissipation due to Eq. \eqref{eq:tau}, and 2) for $q\geq 1/2$ there is no new constraint on $\epsilon$ except for the linearized constraint $0<\epsilon\leq 1$. In other words, the constraint mainly affects the accelerated solutions. Therefore, for $\epsilon$ we consider as a prior the condition \eqref{epsiloncons} which, since the universe must be in a deceleration stage to the past, is enough to fix it at the current time in order to be valid for all redshift.

It is important to mention that, we also constrain the the $\Lambda$CDM model as a further comparison, with a Hubble parameter as a function of the redshift given by
\begin{equation}\label{HLCDM}
    H=H_{0}\sqrt{\Omega_{m,0}\left(1+z\right)^{3}+1-\Omega_{m,0}},
\end{equation}
whose respective parameter space is $\boldsymbol{\theta}=\left(H_{0},\Omega_{m,0}\right)$, and for which we consider the flat priors $0.55<h<0.85$ and $0<\Omega_{m,0}<1$.

\section{\label{sec:results} Results and Discussion}
\begin{figure}
    \centering
    \includegraphics[scale=0.35]{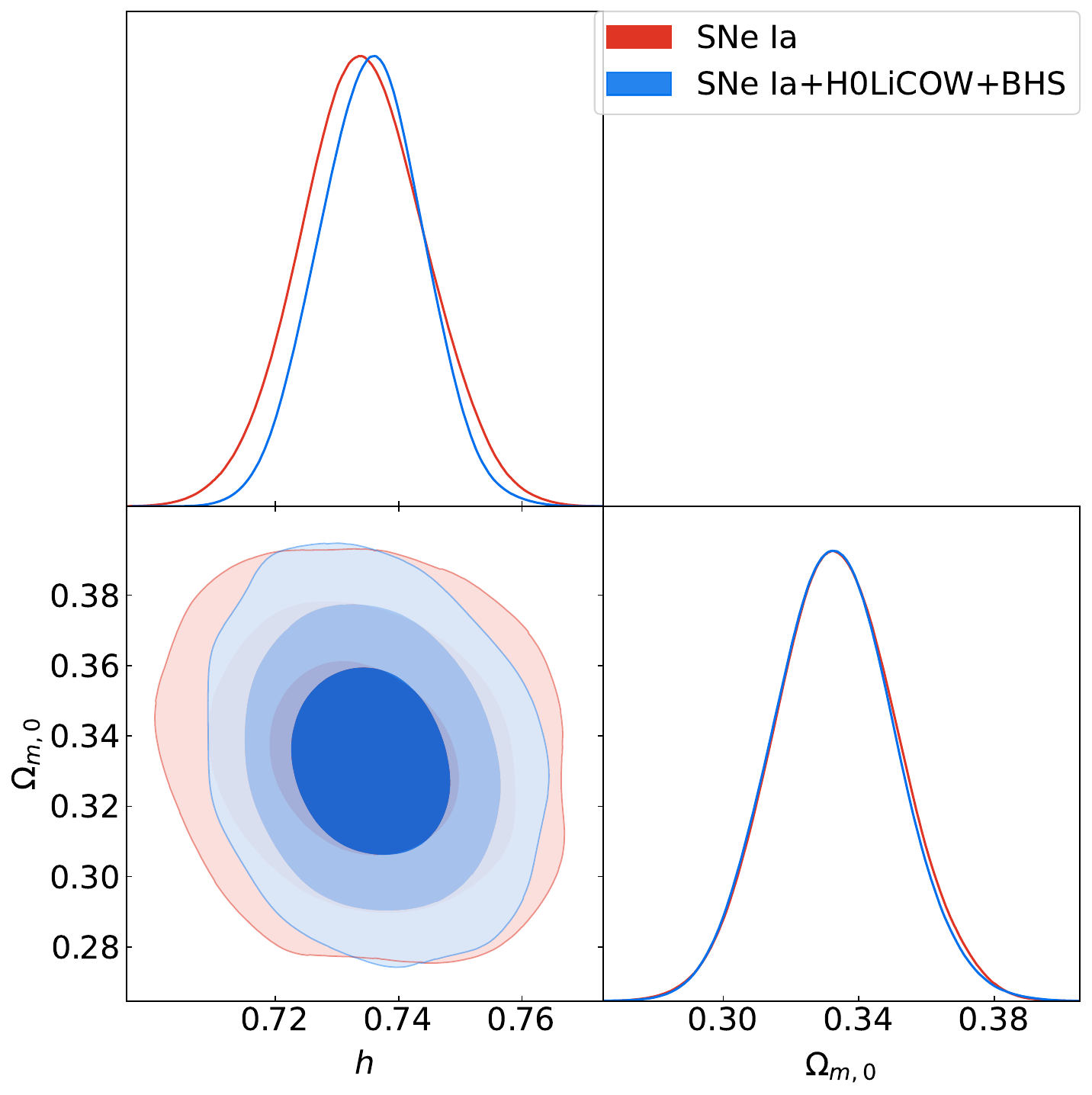}
    \caption{\label{fig:riangleLCDMFit-1} Posterior 1D distribution and joint marginalized regions of the free parameters space of the $\Lambda$CDM model for the SNe Ia and SNe Ia+H0LiCOW+BHS data. The admissible joint regions correspond to $1\sigma$, $2\sigma$, and $3\sigma$ CL, respectively. The best-fit values for each model free parameter are shown in Table \ref{tab:best-fit}.}
\end{figure}

\begin{figure}
    \centering
    \includegraphics[scale=0.35]{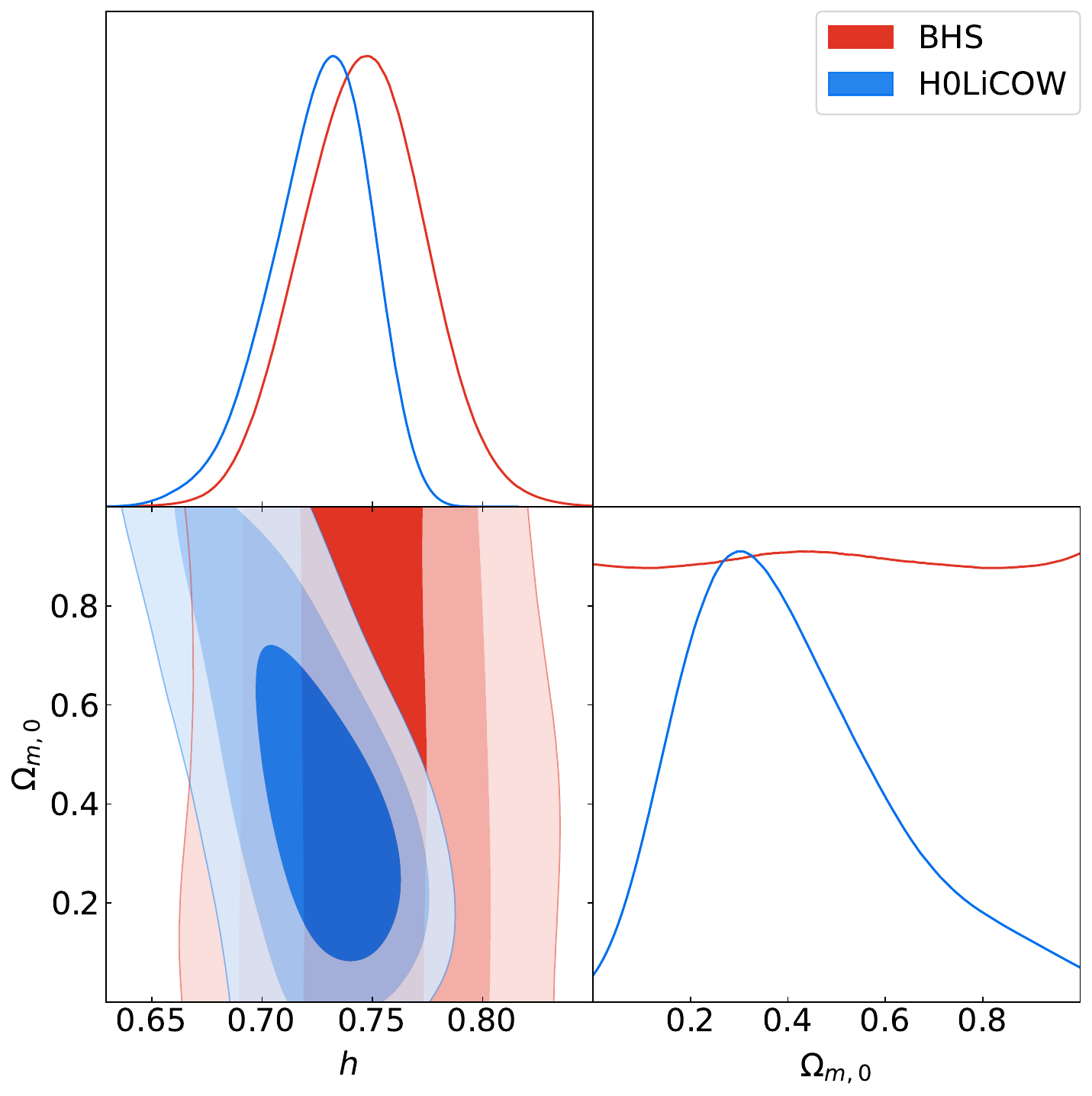}
    \caption{\label{fig:riangleLCDMFit-2} Posterior 1D distribution and joint marginalized regions of the free parameters space of the $\Lambda$CDM model for the BHS and H0LiCOW data. The admissible joint regions correspond to $1\sigma$, $2\sigma$, and $3\sigma$ CL, respectively. The best-fit values for each model free parameter are shown in Table \ref{tab:best-fit}.}
\end{figure}

\begin{figure*}
    \centering
    \includegraphics[scale=0.73]{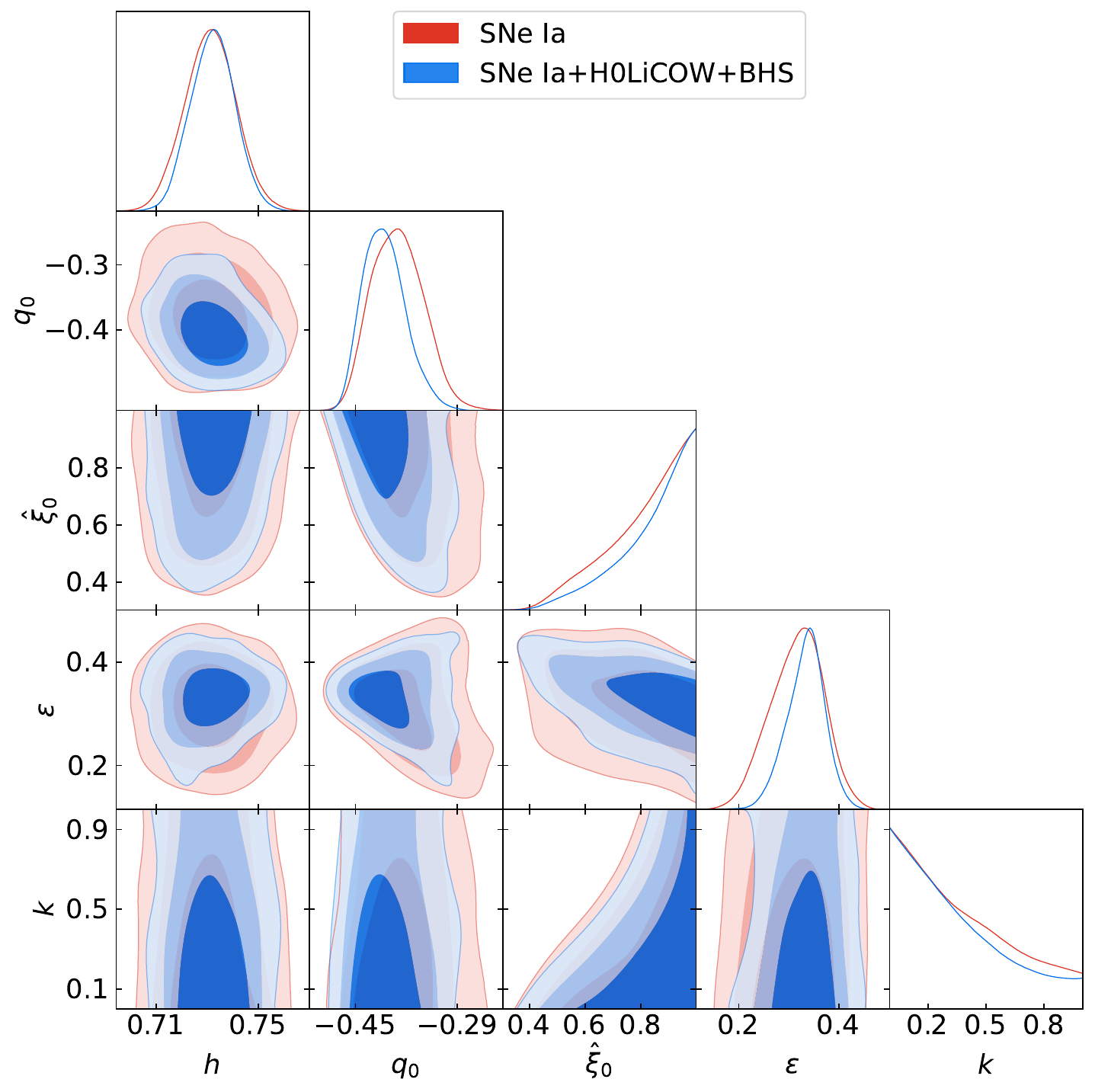}
    \caption{\label{fig:riangleISFit-1} Posterior 1D distribution and joint marginalized regions of the free parameters space of the nonlinear solution of the Israel-Stewart theory for the SNe Ia and SNe Ia+H0LiCOW+BHS data. The admissible joint regions correspond to $1\sigma$, $2\sigma$, and $3\sigma$ CL, respectively. The best-fit values for each model free parameter are shown in Table \ref{tab:best-fit}.}
\end{figure*}

\begin{figure*}
    \centering
    \includegraphics[scale=0.73]{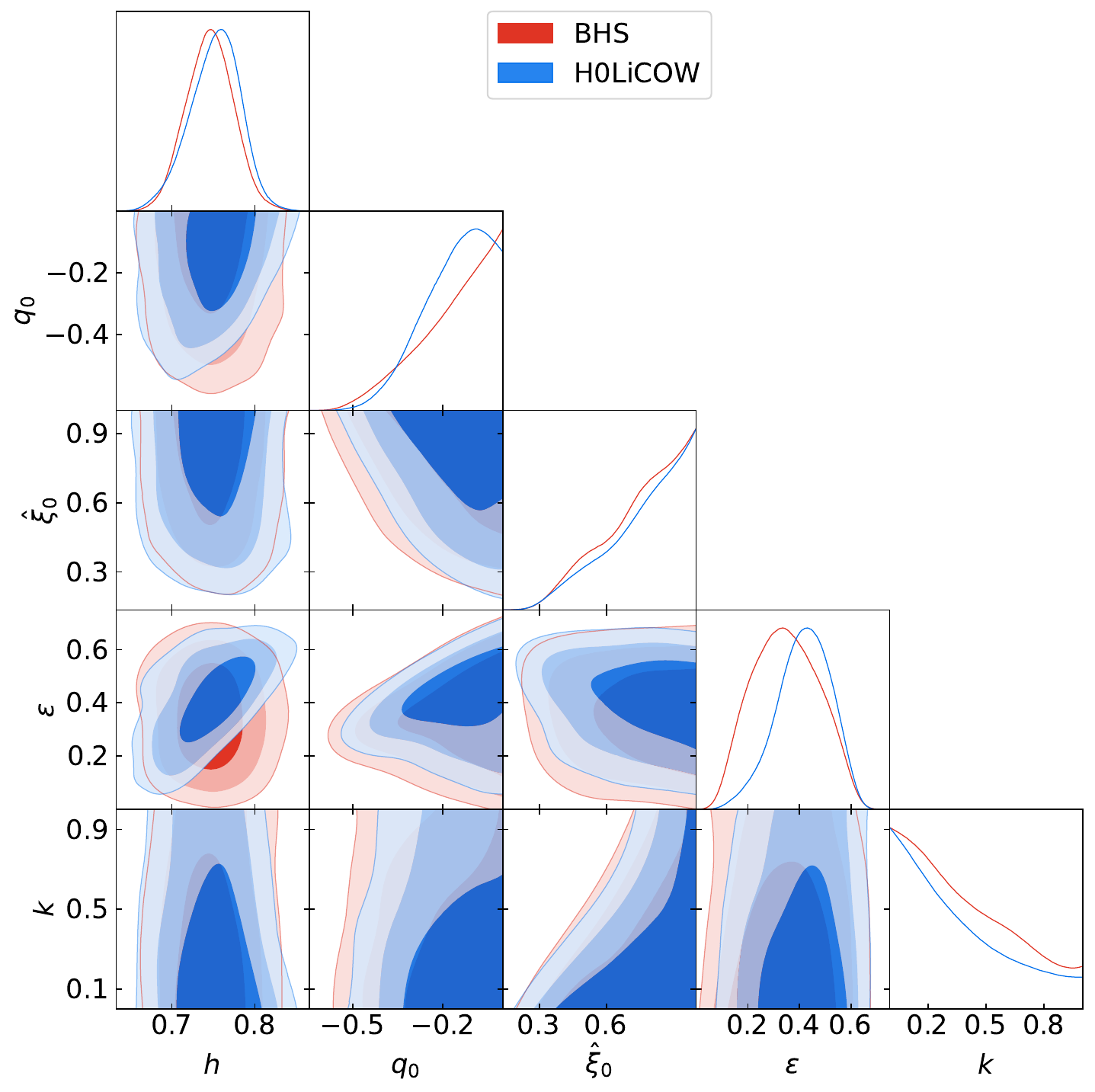}
    \caption{\label{fig:riangleISFit-2} Posterior 1D distribution and joint marginalized regions of the free parameters space of the nonlinear solution of the Israel-Stewart theory for the BHS and H0LiCOW data. The admissible joint regions correspond to $1\sigma$, $2\sigma$, and $3\sigma$ CL, respectively. The best-fit values for each model free parameter are shown in Table \ref{tab:best-fit}.}
\end{figure*}

In Table \ref{tab:best-fit}, we present the best-fit values at $1\sigma$ CL for the free parameters $h$ and $\Omega_{m,0}$ of the $\Lambda$CDM model, and $h$, $q_{0}$, $\hat{\xi}_{0}$, $\epsilon$, $k$ of the nonlinear solution of the Israel-Stewart theory. We also present the $\chi_{\text{min}}^{2}$ criteria obtained for the SNe Ia, H0LiCOW, BHS, and SNe Ia+H0LiCOW+BHS data. In Figs. \ref{fig:riangleLCDMFit-1} and \ref{fig:riangleLCDMFit-2}, we depict the posterior 1D distribution and joint marginalized regions of the free parameters of the $\Lambda$CDM model for the SNe Ia data and the joint analysis, and for the BHS and H0LiCOW data, respectively. On the other hand, In Figs. \ref{fig:riangleISFit-1} and \ref{fig:riangleISFit-2}, we depict the posterior 1D distribution and joint marginalized regions of the free parameters of the nonlinear solution of the Israel-Stewart theory for the SNe Ia data and the joint analysis, and for the BHS and H0LiCOW data, respectively. The admissible joint regions presented in the figures correspond to $1\sigma$, $2\sigma \,(95.5\%)$, and $3\sigma \,(99.7\%)$ CL, respectively. This division between data sets is made for illustrative purposes considering the BHS and H0LiCOW data are only sensitive to $H_{0}$ and, therefore, it is not possible in principle to obtain a best-fit for the other parameters. This last fact translate into a plot that covers all the prior considered for that parameters independently of the CL considered.

\begin{table*}
    \centering
    \begin{tabularx}{\textwidth}{YYYYYYYY}
       \hline\hline
       \multirow{2}{*}{Data} & \multicolumn{6}{c}{Best-fit values} & \multirow{2}{*}{$\chi_{\text{min}}^{2}$} \\
       \cline{2-7}
        & $H_{0}\,\left(\frac{km/s}{Mpc}\right)$ & $\Omega_{m,0}$ & $q_{0}$ & $\hat{\xi}_{0}$ & $\epsilon$ & $k$ &  \\
       \hline
       \multicolumn{8}{c}{$\Lambda$CDM model} \\
       SNe Ia & $73.4_{-1.0}^{+1.0}$ & $0.33_{-0.02}^{+0.02}$ & $\cdots$ & $\cdots$ & $\cdots$ & $\cdots$ & $1523$ \\
       H0LiCOW & $72.8_{-2.5}^{+2.0}$ & $0.37_{-0.18}^{+0.25}$ & $\cdots$ & $\cdots$ & $\cdots$ & $\cdots$ & $132.1$ \\
       BHS & $74.6_{-2.8}^{+2.7}$ & $0.50_{-0.34}^{+0.35}$ & $\cdots$ & $\cdots$ & $\cdots$ & $\cdots$ & $0.865$ \\
       Joint & $73.6_{-0.9}^{+0.8}$ & $0.33_{-0.02}^{+0.02}$ & $\cdots$ & $\cdots$ & $\cdots$ & $\cdots$ & $1656$ \\
       \hline
       \multicolumn{8}{c}{Nonlinear solution} \\
       SNe Ia & $73.1_{-1.0}^{+1.0}$ & $1$ & $-0.38_{-0.04}^{+0.04}$ & $0.85_{-0.19}^{+0.11}$ & $0.32_{-0.06}^{+0.05}$ & $0.31_{-0.23}^{+0.37}$ & $1525$ \\
       H0LiCOW & $75.4_{-3.3}^{+2.8}$ & $1$ & $-0.14_{-0.13}^{+0.10}$ & $0.81_{-0.24}^{+0.14}$ & $0.42_{-0.11}^{+0.10}$ & $0.28_{-0.21}^{+0.38}$ & $130.3$ \\
       BHS & $74.6_{-3.0}^{+2.8}$ & $1$ & $-0.14_{-0.16}^{+0.10}$ & $0.79_{-0.23}^{+0.15}$ & $0.34_{-0.13}^{+0.14}$ & $0.32_{-0.23}^{+0.36}$ & $0.859$ \\
       Joint & $73.2_{-0.9}^{+0.8}$ & $1$ & $-0.41_{-0.03}^{+0.03}$ & $0.88_{-0.17}^{+0.09}$ & $0.34_{-0.04}^{+0.03}$ & $0.27_{-0.20}^{+0.37}$ & $1657$ \\
       \hline\hline
    \end{tabularx}
    \caption{\label{tab:best-fit} Best-fit values and $\chi_{\text{min}}^{2}$ criteria of the $\Lambda$CDM model and the nonlinear solution of the Israel-Stewart theory for the SNe Ia, H0LiCOW, BHS, and SNe Ia+H0LiCOW+BHS data. The uncertainties presented correspond to $1\sigma$ CL.}
\end{table*}

The results presented in Table \ref{tab:best-fit} indicate that the nonlinear solution has slightly lower values of $\chi^{2}_{\text{min}}$ than the $\Lambda$CDM model for the H0LiCOW and BHS data, while this last one has slightly lower values that the nonlinear solution for the SNe Ia data and in the joint analysis. Therefore, considering that the difference in the $\chi^{2}_{\text{min}}$ values are practically negligible, we can conclude that both models are competitive in the description of the SNe Ia, H0LiCOW, and BHS data, including their joint analysis. On physical grounds, the main difference between these models is that in the nonlinear solution, we are describing the recent acceleration expansion of the universe without invoking a cosmological constant, and, at the same time, we are ensuring the causality, local existence, and uniqueness of the solution.

Before discussing the best-fit parameters obtained, we need to highlight an important result. From Fig. \ref{fig:riangleLCDMFit-2} we can see how effectively the BHS and H0LiCOW data are not able to constraint parameters other than $H_{0}$ (in this case, $\Omega_{m,0}$). Both data are considered in this paper to constraint the value of $H_{0}$ in a model-independent way. Nevertheless, the H0LiCOW data is capable of weakly constrain other parameters, as can be seen from the same figure \ref{fig:riangleLCDMFit-2} (in other words, we obtain a best-fit at $1\sigma$ CL). This explains why in Table \ref{tab:best-fit} we present the same best-fit value for $H_{0}$ for the SNe Ia data and the joint analysis, but different best-fit values in the other parameters for the same data. Nevertheless, from Fig. \ref{fig:riangleISFit-2}, we can see that we obtain a constraint on $\epsilon$ and a lower bound for $q_{0}$ and $\hat{\xi}_{0}$ for the BHS and H0LiCOW data. This is a consequence of the condition \eqref{epsiloncons} and the restriction $\kappa\left(\alpha-\beta\right)<q_{0}+1<\kappa\left(\alpha+\beta\right)$ imposed in our constraint on the Eq. \eqref{defofbC} in order to obtain a real Hubble parameter. These values at $3\sigma$ are $0.07<\epsilon<0.64$, $q_{0}>-0.55$, and $\hat{\xi}_{0}>0.27$ for the BHS data, and $0.09<\epsilon<0.65$, $q_{0}>-0.49$, and $\hat{\xi}_{0}>0.26$ for the H0LiCOW data. Therefore, we can consider these intervals as an approximation of the values that must take the parameters $q_{0}$, $\hat{\xi}_{0}$ and $\epsilon$ to obtain a real solution compatible with the condition of causality, local existence, and uniqueness in the nonlinear Israel-Stewart theory.

Focusing on the best-fit values obtained in the joint analysis presented in Table \ref{tab:best-fit}, we can see that the late-time data considered in this paper is not able to strongly constrain the parameter $k$. This conclusion is supported by the 1D distribution and marginalized joint regions presented in Figs. \ref{fig:riangleISFit-1} and \ref{fig:riangleISFit-2}. Nevertheless, we can see through the posterior distribution that a value close to zero is more probably than a value close to $1$, which is compatible with the consideration to obtain this nonlinear solution in which $k^{2}<1$. A similar result is obtained for $\hat{\xi}_{0}$, where we only obtain the lower bound $\hat{\xi}_{0}>0.71$ at $1\sigma$ CL. Considering the change $\xi_{0}=\hat{\xi}_{0}/(1-\hat{\xi}_{0})$, the lower bound obtained translate into $\xi_{0}>2.45$. On the other hand, we obtain strong best-fit on the free parameters $H_{0}$, $q_{0}$, and $\epsilon$, with a value for the first one of $H_{0}=73.6\pm 0.9\,\frac{km/s}{Mpc}$, which is compatible with the value $H_{0}=73.30\pm 1.04\,\frac{km/s}{Mpc}$ obtained by local observations of Cepheids \cite{Riess:2021jrx}. The best-fit obtained for $q_{0}$ is of interest, which with a value of $q_{0}=-0.41\pm 0.03$ at $1\sigma$ CL represent a lower value in comparison with the value $q_{0}=-0.51\pm 0.02$ obtained also by local observations of Cepheids \cite{Riess:2021jrx}. This bet-fit indicates that the nonlinear solution represents a less accelerated universe than the universe represented by the $\Lambda$CDM model. Finally, for the causality parameter, we obtain the best-fit $\epsilon=0.34\pm 0.04$, which is a value strongly incompatible with the condition $\epsilon=1$ considered in many works in the relation between $\tau$ and $\rho$. This highlight the importance of the expression \eqref{eq:tau}, and the conditions \eqref{nonlinearcausality} and \eqref{FOSHcondition} in the Israel-Stewart theory, conditions that ensure the causality of the theory and also leads to solutions that are unique and with local existence. 

\section{\label{sec:conclusions} Conclusions and final remarks}
In this work, we explored the restriction of the cosmological parameters appearing in an exact solution found in the context of a nonlinear regime of the Israel-Stewart theory by using the SNe Ia, H0LiCOW, and BHS data. This nonlinear solution represents a generalization of the one found previously by some of the authors in Ref. \cite{Cruz:2018psw} for the linear version of the Israel-Stewart theory, where the near equilibrium condition is assumed. It is worthy to mention that the exact nonlinear solution considered in our analysis is not the most general. In fact, the differential equation obeyed by the Hubble parameter was simplified by means of the thermodynamic condition of positive entropy production, as discussed in Ref. \cite{Chimento:1997ga}. Through out this work, we set $\omega=0$ in order to compare our results with the $\Lambda$CDM model, and we consider the condition \eqref{epsiloncons}, which ensure the causality, local existence, and uniqueness of the solution.

From the $\chi^{2}_{\text{min}}$ criteria presented in Table \ref{tab:best-fit}, we can conclude that the $\Lambda$CDM model and the nonlinear solution of the Israel-Stewart theory are competitive in the description of the SNe Ia, H0LiCOW, BHS, and SNe Ia+H0LiCOW+BHS data. However, we need to highlight that the nonlinear solution is describing the recent acceleration in the universe expansion without the inclusion of a DE component, on the contrary to the the $\Lambda$CDM model. 

Also, from the the best-fit obtained for the BHS and H0LiCOW data, we can obtain approximate validity intervals for the free parameters $q_{0}$, $\hat{\xi}_{0}$ and $\epsilon$ for which the causality, local existence, and uniqueness of the the nonlinear solution (imposed to be real) is insured, according to the Eq. \eqref{epsiloncons}. These intervals can be condense in only one interval for each free parameter of the form $q_{0}>-0.55$, $\hat{\xi}_{0}>0.26$, and $0.07<\epsilon<0.65$.

Focusing on the best-fit values obtained in the joint analysis, we can see that the the parameter $k$ has a posterior 1D distribution that is close to zero, which is compatible with the consideration to obtain this nonlinear solution where $k^{2}<1$. Also, we obtain the lower bound $\hat{\xi}_{0}>0.71$ at $1\sigma$ CL, which translate into $\xi_{0}>2.45$. This last value can leads to some undesirable physical implications that are not of interest of this paper, but that we plan to address in the future. For the purposes of this paper, the important fact is that this value is compatible with the causality condition of the nonlinear extension of the Israel-Stewart theory. On the other hand, while we obtain a value of $H_{0}$ compatible with the model-independent measurements of Cepheid, we obtain a lower value of $q_{0}$ in comparison with the same measurements. This indicates that the nonlinear solution represents a less accelerated universe than the universe represented by the $\Lambda$CDM model. Finally, the most important result is the best-fit obtained for the causality parameter $\epsilon$, which highlight the importance of the conditions that ensure the causality, local existence, and uniqueness in the nonlinear extension of the Israel-Stewart theory. 

Therefore, we can consider the nonlinear regime of the IS theory a good theoretical framework to describe the recent accelerated expansion of the universe without the inclusion of a cosmological constant or some type of DE. Nevertheless, it remains as an open question if the exploration of viscous fluids in a full nonlinear causal thermodynamics, including more general expressions for the bulk viscosity, could lead to cosmological models more consistent than the $\Lambda$CDM model. We will consider this in the near future.      

\section*{\label{sec:Acknowledgments} Acknowledgments}
M.C. acknowledges the hospitality of the Instituto de Física of Pontificia Universidad Católica de Valparaíso and Departamento de Física of Universidad de Santiago de Chile. E.G. was funded by Vicerrectoría de Investigación y Desarrollo Tecnológico (VRIDT) at Universidad Católica del Norte (UCN) through Proyecto de Investigación Pro Fondecyt 2023, Resolución VRIDT N°076/2023. He also acknowledges the scientific support of Núcleo de Investigación No. 7 UCN-VRIDT 076/2020, Núcleo de Modelación y Simulación Científica (NMSC).

\bibliographystyle{apsrev4-2}
\bibliography{bibliography}

\begin{thebibliography}{62}%
\makeatletter
\providecommand \@ifxundefined [1]{%
 \@ifx{#1\undefined}
}%
\providecommand \@ifnum [1]{%
 \ifnum #1\expandafter \@firstoftwo
 \else \expandafter \@secondoftwo
 \fi
}%
\providecommand \@ifx [1]{%
 \ifx #1\expandafter \@firstoftwo
 \else \expandafter \@secondoftwo
 \fi
}%
\providecommand \natexlab [1]{#1}%
\providecommand \enquote  [1]{``#1''}%
\providecommand \bibnamefont  [1]{#1}%
\providecommand \bibfnamefont [1]{#1}%
\providecommand \citenamefont [1]{#1}%
\providecommand \href@noop [0]{\@secondoftwo}%
\providecommand \href [0]{\begingroup \@sanitize@url \@href}%
\providecommand \@href[1]{\@@startlink{#1}\@@href}%
\providecommand \@@href[1]{\endgroup#1\@@endlink}%
\providecommand \@sanitize@url [0]{\catcode `\\12\catcode `\$12\catcode `\&12\catcode `\#12\catcode `\^12\catcode `\_12\catcode `\%12\relax}%
\providecommand \@@startlink[1]{}%
\providecommand \@@endlink[0]{}%
\providecommand \url  [0]{\begingroup\@sanitize@url \@url }%
\providecommand \@url [1]{\endgroup\@href {#1}{\urlprefix }}%
\providecommand \urlprefix  [0]{URL }%
\providecommand \Eprint [0]{\href }%
\providecommand \doibase [0]{https://doi.org/}%
\providecommand \selectlanguage [0]{\@gobble}%
\providecommand \bibinfo  [0]{\@secondoftwo}%
\providecommand \bibfield  [0]{\@secondoftwo}%
\providecommand \translation [1]{[#1]}%
\providecommand \BibitemOpen [0]{}%
\providecommand \bibitemStop [0]{}%
\providecommand \bibitemNoStop [0]{.\EOS\space}%
\providecommand \EOS [0]{\spacefactor3000\relax}%
\providecommand \BibitemShut  [1]{\csname bibitem#1\endcsname}%
\let\auto@bib@innerbib\@empty
\bibitem [{\citenamefont {Di~Valentino}\ \emph {et~al.}(2019)\citenamefont {Di~Valentino}, \citenamefont {Melchiorri},\ and\ \citenamefont {Silk}}]{DiValentino:2019qzk}%
  \BibitemOpen
  \bibfield  {author} {\bibinfo {author} {\bibfnamefont {E.}~\bibnamefont {Di~Valentino}}, \bibinfo {author} {\bibfnamefont {A.}~\bibnamefont {Melchiorri}},\ and\ \bibinfo {author} {\bibfnamefont {J.}~\bibnamefont {Silk}},\ }\href {https://doi.org/10.1038/s41550-019-0906-9} {\bibfield  {journal} {\bibinfo  {journal} {Nature Astron.}\ }\textbf {\bibinfo {volume} {4}},\ \bibinfo {pages} {196} (\bibinfo {year} {2019})},\ \Eprint {https://arxiv.org/abs/1911.02087} {arXiv:1911.02087 [astro-ph.CO]} \BibitemShut {NoStop}%
\bibitem [{\citenamefont {Handley}(2021)}]{Handley:2019tkm}%
  \BibitemOpen
  \bibfield  {author} {\bibinfo {author} {\bibfnamefont {W.}~\bibnamefont {Handley}},\ }\href {https://doi.org/10.1103/PhysRevD.103.L041301} {\bibfield  {journal} {\bibinfo  {journal} {Phys. Rev. D}\ }\textbf {\bibinfo {volume} {103}},\ \bibinfo {pages} {L041301} (\bibinfo {year} {2021})},\ \Eprint {https://arxiv.org/abs/1908.09139} {arXiv:1908.09139 [astro-ph.CO]} \BibitemShut {NoStop}%
\bibitem [{\citenamefont {Guth}(1981)}]{Guth:1980zm}%
  \BibitemOpen
  \bibfield  {author} {\bibinfo {author} {\bibfnamefont {A.~H.}\ \bibnamefont {Guth}},\ }\href {https://doi.org/10.1103/PhysRevD.23.347} {\bibfield  {journal} {\bibinfo  {journal} {Phys. Rev. D}\ }\textbf {\bibinfo {volume} {23}},\ \bibinfo {pages} {347} (\bibinfo {year} {1981})}\BibitemShut {NoStop}%
\bibitem [{\citenamefont {Weinberg}(1989)}]{Weinberg:1988cp}%
  \BibitemOpen
  \bibfield  {author} {\bibinfo {author} {\bibfnamefont {S.}~\bibnamefont {Weinberg}},\ }\href {https://doi.org/10.1103/RevModPhys.61.1} {\bibfield  {journal} {\bibinfo  {journal} {Rev. Mod. Phys.}\ }\textbf {\bibinfo {volume} {61}},\ \bibinfo {pages} {1} (\bibinfo {year} {1989})}\BibitemShut {NoStop}%
\bibitem [{\citenamefont {Padmanabhan}(2003)}]{Padmanabhan:2002ji}%
  \BibitemOpen
  \bibfield  {author} {\bibinfo {author} {\bibfnamefont {T.}~\bibnamefont {Padmanabhan}},\ }\href {https://doi.org/10.1016/S0370-1573(03)00120-0} {\bibfield  {journal} {\bibinfo  {journal} {Phys. Rept.}\ }\textbf {\bibinfo {volume} {380}},\ \bibinfo {pages} {235} (\bibinfo {year} {2003})},\ \Eprint {https://arxiv.org/abs/hep-th/0212290} {arXiv:hep-th/0212290} \BibitemShut {NoStop}%
\bibitem [{\citenamefont {Velten}\ \emph {et~al.}(2014)\citenamefont {Velten}, \citenamefont {vom Marttens},\ and\ \citenamefont {Zimdahl}}]{Velten:2014nra}%
  \BibitemOpen
  \bibfield  {author} {\bibinfo {author} {\bibfnamefont {H.~E.~S.}\ \bibnamefont {Velten}}, \bibinfo {author} {\bibfnamefont {R.~F.}\ \bibnamefont {vom Marttens}},\ and\ \bibinfo {author} {\bibfnamefont {W.}~\bibnamefont {Zimdahl}},\ }\href {https://doi.org/10.1140/epjc/s10052-014-3160-4} {\bibfield  {journal} {\bibinfo  {journal} {Eur. Phys. J. C}\ }\textbf {\bibinfo {volume} {74}},\ \bibinfo {pages} {3160} (\bibinfo {year} {2014})},\ \Eprint {https://arxiv.org/abs/1410.2509} {arXiv:1410.2509 [astro-ph.CO]} \BibitemShut {NoStop}%
\bibitem [{\citenamefont {Riess}\ \emph {et~al.}(2022)\citenamefont {Riess} \emph {et~al.}}]{Riess:2021jrx}%
  \BibitemOpen
  \bibfield  {author} {\bibinfo {author} {\bibfnamefont {A.~G.}\ \bibnamefont {Riess}} \emph {et~al.},\ }\href {https://doi.org/10.3847/2041-8213/ac5c5b} {\bibfield  {journal} {\bibinfo  {journal} {Astrophys. J. Lett.}\ }\textbf {\bibinfo {volume} {934}},\ \bibinfo {pages} {L7} (\bibinfo {year} {2022})},\ \Eprint {https://arxiv.org/abs/2112.04510} {arXiv:2112.04510 [astro-ph.CO]} \BibitemShut {NoStop}%
\bibitem [{\citenamefont {Wong}\ \emph {et~al.}(2020)\citenamefont {Wong} \emph {et~al.}}]{Wong:2019kwg}%
  \BibitemOpen
  \bibfield  {author} {\bibinfo {author} {\bibfnamefont {K.~C.}\ \bibnamefont {Wong}} \emph {et~al.},\ }\href {https://doi.org/10.1093/mnras/stz3094} {\bibfield  {journal} {\bibinfo  {journal} {Mon. Not. Roy. Astron. Soc.}\ }\textbf {\bibinfo {volume} {498}},\ \bibinfo {pages} {1420} (\bibinfo {year} {2020})},\ \Eprint {https://arxiv.org/abs/1907.04869} {arXiv:1907.04869 [astro-ph.CO]} \BibitemShut {NoStop}%
\bibitem [{\citenamefont {Jaime}\ \emph {et~al.}(2018)\citenamefont {Jaime}, \citenamefont {Jaber},\ and\ \citenamefont {Escamilla-Rivera}}]{Jaime:2018ftn}%
  \BibitemOpen
  \bibfield  {author} {\bibinfo {author} {\bibfnamefont {L.~G.}\ \bibnamefont {Jaime}}, \bibinfo {author} {\bibfnamefont {M.}~\bibnamefont {Jaber}},\ and\ \bibinfo {author} {\bibfnamefont {C.}~\bibnamefont {Escamilla-Rivera}},\ }\href {https://doi.org/10.1103/PhysRevD.98.083530} {\bibfield  {journal} {\bibinfo  {journal} {Phys. Rev. D}\ }\textbf {\bibinfo {volume} {98}},\ \bibinfo {pages} {083530} (\bibinfo {year} {2018})},\ \Eprint {https://arxiv.org/abs/1804.04284} {arXiv:1804.04284 [astro-ph.CO]} \BibitemShut {NoStop}%
\bibitem [{\citenamefont {Jaber}\ \emph {et~al.}(2022)\citenamefont {Jaber}, \citenamefont {Arciniega}, \citenamefont {Jaime},\ and\ \citenamefont {Rodr\'\i{}guez-L\'opez}}]{Jaber:2021hho}%
  \BibitemOpen
  \bibfield  {author} {\bibinfo {author} {\bibfnamefont {M.}~\bibnamefont {Jaber}}, \bibinfo {author} {\bibfnamefont {G.}~\bibnamefont {Arciniega}}, \bibinfo {author} {\bibfnamefont {L.~G.}\ \bibnamefont {Jaime}},\ and\ \bibinfo {author} {\bibfnamefont {O.~A.}\ \bibnamefont {Rodr\'\i{}guez-L\'opez}},\ }\href {https://doi.org/10.1016/j.dark.2022.101069} {\bibfield  {journal} {\bibinfo  {journal} {Phys. Dark Univ.}\ }\textbf {\bibinfo {volume} {37}},\ \bibinfo {pages} {101069} (\bibinfo {year} {2022})},\ \Eprint {https://arxiv.org/abs/2102.08561} {arXiv:2102.08561 [astro-ph.CO]} \BibitemShut {NoStop}%
\bibitem [{\citenamefont {Aghanim}\ \emph {et~al.}(2020)\citenamefont {Aghanim} \emph {et~al.}}]{Planck:2018vyg}%
  \BibitemOpen
  \bibfield  {author} {\bibinfo {author} {\bibfnamefont {N.}~\bibnamefont {Aghanim}} \emph {et~al.} (\bibinfo {collaboration} {Planck}),\ }\href {https://doi.org/10.1051/0004-6361/201833910} {\bibfield  {journal} {\bibinfo  {journal} {Astron. Astrophys.}\ }\textbf {\bibinfo {volume} {641}},\ \bibinfo {pages} {A6} (\bibinfo {year} {2020})},\ \bibinfo {note} {[Erratum: Astron.Astrophys. 652, C4 (2021)]},\ \Eprint {https://arxiv.org/abs/1807.06209} {arXiv:1807.06209 [astro-ph.CO]} \BibitemShut {NoStop}%
\bibitem [{\citenamefont {Israel}(1976)}]{Israel:1976tn}%
  \BibitemOpen
  \bibfield  {author} {\bibinfo {author} {\bibfnamefont {W.}~\bibnamefont {Israel}},\ }\href {https://doi.org/10.1016/0003-4916(76)90064-6} {\bibfield  {journal} {\bibinfo  {journal} {Annals Phys.}\ }\textbf {\bibinfo {volume} {100}},\ \bibinfo {pages} {310} (\bibinfo {year} {1976})}\BibitemShut {NoStop}%
\bibitem [{\citenamefont {Israel}\ and\ \citenamefont {Stewart}(1979)}]{Israel:1979wp}%
  \BibitemOpen
  \bibfield  {author} {\bibinfo {author} {\bibfnamefont {W.}~\bibnamefont {Israel}}\ and\ \bibinfo {author} {\bibfnamefont {J.~M.}\ \bibnamefont {Stewart}},\ }\href {https://doi.org/10.1016/0003-4916(79)90130-1} {\bibfield  {journal} {\bibinfo  {journal} {Annals Phys.}\ }\textbf {\bibinfo {volume} {118}},\ \bibinfo {pages} {341} (\bibinfo {year} {1979})}\BibitemShut {NoStop}%
\bibitem [{\citenamefont {Pavon}(1990)}]{Pavon:1990ha}%
  \BibitemOpen
  \bibfield  {author} {\bibinfo {author} {\bibfnamefont {D.}~\bibnamefont {Pavon}},\ }\href {https://doi.org/10.1088/0264-9381/7/3/022} {\bibfield  {journal} {\bibinfo  {journal} {Class. Quant. Grav.}\ }\textbf {\bibinfo {volume} {7}},\ \bibinfo {pages} {487} (\bibinfo {year} {1990})}\BibitemShut {NoStop}%
\bibitem [{\citenamefont {Chimento}\ and\ \citenamefont {Jakubi}(1993)}]{Chimento:1993zc}%
  \BibitemOpen
  \bibfield  {author} {\bibinfo {author} {\bibfnamefont {L.}~\bibnamefont {Chimento}}\ and\ \bibinfo {author} {\bibfnamefont {A.~S.}\ \bibnamefont {Jakubi}},\ }\href {https://doi.org/10.1088/0264-9381/10/10/011} {\bibfield  {journal} {\bibinfo  {journal} {Class. Quant. Grav.}\ }\textbf {\bibinfo {volume} {10}},\ \bibinfo {pages} {2047} (\bibinfo {year} {1993})}\BibitemShut {NoStop}%
\bibitem [{\citenamefont {Maartens}(1996)}]{Maartens:1996vi}%
  \BibitemOpen
  \bibfield  {author} {\bibinfo {author} {\bibfnamefont {R.}~\bibnamefont {Maartens}}\ }(\bibinfo {year} {1996})\ \Eprint {https://arxiv.org/abs/astro-ph/9609119} {arXiv:astro-ph/9609119} \BibitemShut {NoStop}%
\bibitem [{\citenamefont {Zimdahl}(1996)}]{Zimdahl:1996fj}%
  \BibitemOpen
  \bibfield  {author} {\bibinfo {author} {\bibfnamefont {W.}~\bibnamefont {Zimdahl}},\ }\href {https://doi.org/10.1093/mnras/280.4.1239} {\bibfield  {journal} {\bibinfo  {journal} {Mon. Not. Roy. Astron. Soc.}\ }\textbf {\bibinfo {volume} {280}},\ \bibinfo {pages} {1239} (\bibinfo {year} {1996})},\ \Eprint {https://arxiv.org/abs/astro-ph/9602128} {arXiv:astro-ph/9602128} \BibitemShut {NoStop}%
\bibitem [{\citenamefont {Wilson}\ \emph {et~al.}(2007)\citenamefont {Wilson}, \citenamefont {Mathews},\ and\ \citenamefont {Fuller}}]{Wilson:2006gf}%
  \BibitemOpen
  \bibfield  {author} {\bibinfo {author} {\bibfnamefont {J.~R.}\ \bibnamefont {Wilson}}, \bibinfo {author} {\bibfnamefont {G.~J.}\ \bibnamefont {Mathews}},\ and\ \bibinfo {author} {\bibfnamefont {G.~M.}\ \bibnamefont {Fuller}},\ }\href {https://doi.org/10.1103/PhysRevD.75.043521} {\bibfield  {journal} {\bibinfo  {journal} {Phys. Rev. D}\ }\textbf {\bibinfo {volume} {75}},\ \bibinfo {pages} {043521} (\bibinfo {year} {2007})},\ \Eprint {https://arxiv.org/abs/astro-ph/0609687} {arXiv:astro-ph/0609687} \BibitemShut {NoStop}%
\bibitem [{\citenamefont {Mathews}\ \emph {et~al.}(2008)\citenamefont {Mathews}, \citenamefont {Lan},\ and\ \citenamefont {Kolda}}]{Mathews:2008hk}%
  \BibitemOpen
  \bibfield  {author} {\bibinfo {author} {\bibfnamefont {G.~J.}\ \bibnamefont {Mathews}}, \bibinfo {author} {\bibfnamefont {N.~Q.}\ \bibnamefont {Lan}},\ and\ \bibinfo {author} {\bibfnamefont {C.}~\bibnamefont {Kolda}},\ }\href {https://doi.org/10.1103/PhysRevD.78.043525} {\bibfield  {journal} {\bibinfo  {journal} {Phys. Rev. D}\ }\textbf {\bibinfo {volume} {78}},\ \bibinfo {pages} {043525} (\bibinfo {year} {2008})},\ \Eprint {https://arxiv.org/abs/0801.0853} {arXiv:0801.0853 [astro-ph]} \BibitemShut {NoStop}%
\bibitem [{\citenamefont {Pandey}\ \emph {et~al.}(2020)\citenamefont {Pandey}, \citenamefont {Karwal},\ and\ \citenamefont {Das}}]{Pandey:2019plg}%
  \BibitemOpen
  \bibfield  {author} {\bibinfo {author} {\bibfnamefont {K.~L.}\ \bibnamefont {Pandey}}, \bibinfo {author} {\bibfnamefont {T.}~\bibnamefont {Karwal}},\ and\ \bibinfo {author} {\bibfnamefont {S.}~\bibnamefont {Das}},\ }\href {https://doi.org/10.1088/1475-7516/2020/07/026} {\bibfield  {journal} {\bibinfo  {journal} {JCAP}\ }\textbf {\bibinfo {volume} {07}},\ \bibinfo {pages} {026}},\ \Eprint {https://arxiv.org/abs/1902.10636} {arXiv:1902.10636 [astro-ph.CO]} \BibitemShut {NoStop}%
\bibitem [{\citenamefont {C\'ardenas}\ \emph {et~al.}(2019)\citenamefont {C\'ardenas}, \citenamefont {Grand\'on},\ and\ \citenamefont {Lepe}}]{Cardenas:2018nem}%
  \BibitemOpen
  \bibfield  {author} {\bibinfo {author} {\bibfnamefont {V.~H.}\ \bibnamefont {C\'ardenas}}, \bibinfo {author} {\bibfnamefont {D.}~\bibnamefont {Grand\'on}},\ and\ \bibinfo {author} {\bibfnamefont {S.}~\bibnamefont {Lepe}},\ }\href {https://doi.org/10.1140/epjc/s10052-019-6887-0} {\bibfield  {journal} {\bibinfo  {journal} {Eur. Phys. J. C}\ }\textbf {\bibinfo {volume} {79}},\ \bibinfo {pages} {357} (\bibinfo {year} {2019})},\ \Eprint {https://arxiv.org/abs/1812.03540} {arXiv:1812.03540 [astro-ph.CO]} \BibitemShut {NoStop}%
\bibitem [{\citenamefont {Wang}\ \emph {et~al.}(2016)\citenamefont {Wang}, \citenamefont {Abdalla}, \citenamefont {Atrio-Barandela},\ and\ \citenamefont {Pavon}}]{Wang:2016lxa}%
  \BibitemOpen
  \bibfield  {author} {\bibinfo {author} {\bibfnamefont {B.}~\bibnamefont {Wang}}, \bibinfo {author} {\bibfnamefont {E.}~\bibnamefont {Abdalla}}, \bibinfo {author} {\bibfnamefont {F.}~\bibnamefont {Atrio-Barandela}},\ and\ \bibinfo {author} {\bibfnamefont {D.}~\bibnamefont {Pavon}},\ }\href {https://doi.org/10.1088/0034-4885/79/9/096901} {\bibfield  {journal} {\bibinfo  {journal} {Rept. Prog. Phys.}\ }\textbf {\bibinfo {volume} {79}},\ \bibinfo {pages} {096901} (\bibinfo {year} {2016})},\ \Eprint {https://arxiv.org/abs/1603.08299} {arXiv:1603.08299 [astro-ph.CO]} \BibitemShut {NoStop}%
\bibitem [{\citenamefont {Cataldo}\ \emph {et~al.}(2005)\citenamefont {Cataldo}, \citenamefont {Cruz},\ and\ \citenamefont {Lepe}}]{Cataldo:2005qh}%
  \BibitemOpen
  \bibfield  {author} {\bibinfo {author} {\bibfnamefont {M.}~\bibnamefont {Cataldo}}, \bibinfo {author} {\bibfnamefont {N.}~\bibnamefont {Cruz}},\ and\ \bibinfo {author} {\bibfnamefont {S.}~\bibnamefont {Lepe}},\ }\href {https://doi.org/10.1016/j.physletb.2005.05.029} {\bibfield  {journal} {\bibinfo  {journal} {Phys. Lett. B}\ }\textbf {\bibinfo {volume} {619}},\ \bibinfo {pages} {5} (\bibinfo {year} {2005})},\ \Eprint {https://arxiv.org/abs/hep-th/0506153} {arXiv:hep-th/0506153} \BibitemShut {NoStop}%
\bibitem [{\citenamefont {Disconzi}\ \emph {et~al.}(2015)\citenamefont {Disconzi}, \citenamefont {Kephart},\ and\ \citenamefont {Scherrer}}]{Disconzi:2014oda}%
  \BibitemOpen
  \bibfield  {author} {\bibinfo {author} {\bibfnamefont {M.~M.}\ \bibnamefont {Disconzi}}, \bibinfo {author} {\bibfnamefont {T.~W.}\ \bibnamefont {Kephart}},\ and\ \bibinfo {author} {\bibfnamefont {R.~J.}\ \bibnamefont {Scherrer}},\ }\href {https://doi.org/10.1103/PhysRevD.91.043532} {\bibfield  {journal} {\bibinfo  {journal} {Phys. Rev. D}\ }\textbf {\bibinfo {volume} {91}},\ \bibinfo {pages} {043532} (\bibinfo {year} {2015})},\ \Eprint {https://arxiv.org/abs/1409.4918} {arXiv:1409.4918 [gr-qc]} \BibitemShut {NoStop}%
\bibitem [{\citenamefont {Cruz}\ \emph {et~al.}(2017{\natexlab{a}})\citenamefont {Cruz}, \citenamefont {Cruz},\ and\ \citenamefont {Lepe}}]{Cruz:2017lbu}%
  \BibitemOpen
  \bibfield  {author} {\bibinfo {author} {\bibfnamefont {M.}~\bibnamefont {Cruz}}, \bibinfo {author} {\bibfnamefont {N.}~\bibnamefont {Cruz}},\ and\ \bibinfo {author} {\bibfnamefont {S.}~\bibnamefont {Lepe}},\ }\href {https://doi.org/10.1016/j.physletb.2017.03.065} {\bibfield  {journal} {\bibinfo  {journal} {Phys. Lett. B}\ }\textbf {\bibinfo {volume} {769}},\ \bibinfo {pages} {159} (\bibinfo {year} {2017}{\natexlab{a}})},\ \Eprint {https://arxiv.org/abs/1701.06724} {arXiv:1701.06724 [gr-qc]} \BibitemShut {NoStop}%
\bibitem [{\citenamefont {Cruz}\ \emph {et~al.}(2017{\natexlab{b}})\citenamefont {Cruz}, \citenamefont {Lepe},\ and\ \citenamefont {Pe\~na}}]{Cruz:2016rqi}%
  \BibitemOpen
  \bibfield  {author} {\bibinfo {author} {\bibfnamefont {N.}~\bibnamefont {Cruz}}, \bibinfo {author} {\bibfnamefont {S.}~\bibnamefont {Lepe}},\ and\ \bibinfo {author} {\bibfnamefont {F.}~\bibnamefont {Pe\~na}},\ }\href {https://doi.org/10.1016/j.physletb.2017.01.035} {\bibfield  {journal} {\bibinfo  {journal} {Phys. Lett. B}\ }\textbf {\bibinfo {volume} {767}},\ \bibinfo {pages} {103} (\bibinfo {year} {2017}{\natexlab{b}})},\ \Eprint {https://arxiv.org/abs/1607.04192} {arXiv:1607.04192 [gr-qc]} \BibitemShut {NoStop}%
\bibitem [{\citenamefont {C\'ardenas}\ \emph {et~al.}(2020)\citenamefont {C\'ardenas}, \citenamefont {Cruz},\ and\ \citenamefont {Lepe}}]{Cardenas:2020exv}%
  \BibitemOpen
  \bibfield  {author} {\bibinfo {author} {\bibfnamefont {V.~H.}\ \bibnamefont {C\'ardenas}}, \bibinfo {author} {\bibfnamefont {M.}~\bibnamefont {Cruz}},\ and\ \bibinfo {author} {\bibfnamefont {S.}~\bibnamefont {Lepe}},\ }\href {https://doi.org/10.1103/PhysRevD.102.123543} {\bibfield  {journal} {\bibinfo  {journal} {Phys. Rev. D}\ }\textbf {\bibinfo {volume} {102}},\ \bibinfo {pages} {123543} (\bibinfo {year} {2020})},\ \Eprint {https://arxiv.org/abs/2008.12403} {arXiv:2008.12403 [gr-qc]} \BibitemShut {NoStop}%
\bibitem [{\citenamefont {Cruz}\ \emph {et~al.}(2018{\natexlab{a}})\citenamefont {Cruz}, \citenamefont {Lepe},\ and\ \citenamefont {Odintsov}}]{Cruz:2018arw}%
  \BibitemOpen
  \bibfield  {author} {\bibinfo {author} {\bibfnamefont {M.}~\bibnamefont {Cruz}}, \bibinfo {author} {\bibfnamefont {S.}~\bibnamefont {Lepe}},\ and\ \bibinfo {author} {\bibfnamefont {S.~D.}\ \bibnamefont {Odintsov}},\ }\href {https://doi.org/10.1103/PhysRevD.98.083515} {\bibfield  {journal} {\bibinfo  {journal} {Phys. Rev. D}\ }\textbf {\bibinfo {volume} {98}},\ \bibinfo {pages} {083515} (\bibinfo {year} {2018}{\natexlab{a}})},\ \Eprint {https://arxiv.org/abs/1808.03825} {arXiv:1808.03825 [gr-qc]} \BibitemShut {NoStop}%
\bibitem [{\citenamefont {Bemfica}\ \emph {et~al.}(2019)\citenamefont {Bemfica}, \citenamefont {Disconzi},\ and\ \citenamefont {Noronha}}]{Bemfica:2019cop}%
  \BibitemOpen
  \bibfield  {author} {\bibinfo {author} {\bibfnamefont {F.~S.}\ \bibnamefont {Bemfica}}, \bibinfo {author} {\bibfnamefont {M.~M.}\ \bibnamefont {Disconzi}},\ and\ \bibinfo {author} {\bibfnamefont {J.}~\bibnamefont {Noronha}},\ }\href {https://doi.org/10.1103/PhysRevLett.122.221602} {\bibfield  {journal} {\bibinfo  {journal} {Phys. Rev. Lett.}\ }\textbf {\bibinfo {volume} {122}},\ \bibinfo {pages} {221602} (\bibinfo {year} {2019})},\ \Eprint {https://arxiv.org/abs/1901.06701} {arXiv:1901.06701 [gr-qc]} \BibitemShut {NoStop}%
\bibitem [{\citenamefont {Alford}\ \emph {et~al.}(2018)\citenamefont {Alford}, \citenamefont {Bovard}, \citenamefont {Hanauske}, \citenamefont {Rezzolla},\ and\ \citenamefont {Schwenzer}}]{Alford:2017rxf}%
  \BibitemOpen
  \bibfield  {author} {\bibinfo {author} {\bibfnamefont {M.~G.}\ \bibnamefont {Alford}}, \bibinfo {author} {\bibfnamefont {L.}~\bibnamefont {Bovard}}, \bibinfo {author} {\bibfnamefont {M.}~\bibnamefont {Hanauske}}, \bibinfo {author} {\bibfnamefont {L.}~\bibnamefont {Rezzolla}},\ and\ \bibinfo {author} {\bibfnamefont {K.}~\bibnamefont {Schwenzer}},\ }\href {https://doi.org/10.1103/PhysRevLett.120.041101} {\bibfield  {journal} {\bibinfo  {journal} {Phys. Rev. Lett.}\ }\textbf {\bibinfo {volume} {120}},\ \bibinfo {pages} {041101} (\bibinfo {year} {2018})},\ \Eprint {https://arxiv.org/abs/1707.09475} {arXiv:1707.09475 [gr-qc]} \BibitemShut {NoStop}%
\bibitem [{\citenamefont {Barta}(2019)}]{Barta:2019tpv}%
  \BibitemOpen
  \bibfield  {author} {\bibinfo {author} {\bibfnamefont {D.}~\bibnamefont {Barta}},\ }\href {https://doi.org/10.1088/1361-6382/ab472e} {\bibfield  {journal} {\bibinfo  {journal} {Class. Quant. Grav.}\ }\textbf {\bibinfo {volume} {36}},\ \bibinfo {pages} {215012} (\bibinfo {year} {2019})},\ \Eprint {https://arxiv.org/abs/1904.00907} {arXiv:1904.00907 [gr-qc]} \BibitemShut {NoStop}%
\bibitem [{\citenamefont {Bravo~Medina}\ \emph {et~al.}(2019)\citenamefont {Bravo~Medina}, \citenamefont {Nowakowski},\ and\ \citenamefont {Batic}}]{BravoMedina:2019han}%
  \BibitemOpen
  \bibfield  {author} {\bibinfo {author} {\bibfnamefont {S.}~\bibnamefont {Bravo~Medina}}, \bibinfo {author} {\bibfnamefont {M.}~\bibnamefont {Nowakowski}},\ and\ \bibinfo {author} {\bibfnamefont {D.}~\bibnamefont {Batic}},\ }\href {https://doi.org/10.1088/1361-6382/ab45bb} {\bibfield  {journal} {\bibinfo  {journal} {Class. Quant. Grav.}\ }\textbf {\bibinfo {volume} {36}},\ \bibinfo {pages} {215002} (\bibinfo {year} {2019})},\ \Eprint {https://arxiv.org/abs/1901.09787} {arXiv:1901.09787 [gr-qc]} \BibitemShut {NoStop}%
\bibitem [{\citenamefont {Yang}\ \emph {et~al.}(2019)\citenamefont {Yang}, \citenamefont {Pan}, \citenamefont {Di~Valentino}, \citenamefont {Paliathanasis},\ and\ \citenamefont {Lu}}]{Yang:2019qza}%
  \BibitemOpen
  \bibfield  {author} {\bibinfo {author} {\bibfnamefont {W.}~\bibnamefont {Yang}}, \bibinfo {author} {\bibfnamefont {S.}~\bibnamefont {Pan}}, \bibinfo {author} {\bibfnamefont {E.}~\bibnamefont {Di~Valentino}}, \bibinfo {author} {\bibfnamefont {A.}~\bibnamefont {Paliathanasis}},\ and\ \bibinfo {author} {\bibfnamefont {J.}~\bibnamefont {Lu}},\ }\href {https://doi.org/10.1103/PhysRevD.100.103518} {\bibfield  {journal} {\bibinfo  {journal} {Phys. Rev. D}\ }\textbf {\bibinfo {volume} {100}},\ \bibinfo {pages} {103518} (\bibinfo {year} {2019})},\ \Eprint {https://arxiv.org/abs/1906.04162} {arXiv:1906.04162 [astro-ph.CO]} \BibitemShut {NoStop}%
\bibitem [{\citenamefont {Poulin}\ \emph {et~al.}(2019)\citenamefont {Poulin}, \citenamefont {Smith}, \citenamefont {Karwal},\ and\ \citenamefont {Kamionkowski}}]{Poulin:2018cxd}%
  \BibitemOpen
  \bibfield  {author} {\bibinfo {author} {\bibfnamefont {V.}~\bibnamefont {Poulin}}, \bibinfo {author} {\bibfnamefont {T.~L.}\ \bibnamefont {Smith}}, \bibinfo {author} {\bibfnamefont {T.}~\bibnamefont {Karwal}},\ and\ \bibinfo {author} {\bibfnamefont {M.}~\bibnamefont {Kamionkowski}},\ }\href {https://doi.org/10.1103/PhysRevLett.122.221301} {\bibfield  {journal} {\bibinfo  {journal} {Phys. Rev. Lett.}\ }\textbf {\bibinfo {volume} {122}},\ \bibinfo {pages} {221301} (\bibinfo {year} {2019})},\ \Eprint {https://arxiv.org/abs/1811.04083} {arXiv:1811.04083 [astro-ph.CO]} \BibitemShut {NoStop}%
\bibitem [{\citenamefont {Di~Valentino}\ \emph {et~al.}(2021)\citenamefont {Di~Valentino}, \citenamefont {Mena}, \citenamefont {Pan}, \citenamefont {Visinelli}, \citenamefont {Yang}, \citenamefont {Melchiorri}, \citenamefont {Mota}, \citenamefont {Riess},\ and\ \citenamefont {Silk}}]{DiValentino:2021izs}%
  \BibitemOpen
  \bibfield  {author} {\bibinfo {author} {\bibfnamefont {E.}~\bibnamefont {Di~Valentino}}, \bibinfo {author} {\bibfnamefont {O.}~\bibnamefont {Mena}}, \bibinfo {author} {\bibfnamefont {S.}~\bibnamefont {Pan}}, \bibinfo {author} {\bibfnamefont {L.}~\bibnamefont {Visinelli}}, \bibinfo {author} {\bibfnamefont {W.}~\bibnamefont {Yang}}, \bibinfo {author} {\bibfnamefont {A.}~\bibnamefont {Melchiorri}}, \bibinfo {author} {\bibfnamefont {D.~F.}\ \bibnamefont {Mota}}, \bibinfo {author} {\bibfnamefont {A.~G.}\ \bibnamefont {Riess}},\ and\ \bibinfo {author} {\bibfnamefont {J.}~\bibnamefont {Silk}},\ }\href {https://doi.org/10.1088/1361-6382/ac086d} {\bibfield  {journal} {\bibinfo  {journal} {Class. Quant. Grav.}\ }\textbf {\bibinfo {volume} {38}},\ \bibinfo {pages} {153001} (\bibinfo {year} {2021})},\ \Eprint {https://arxiv.org/abs/2103.01183} {arXiv:2103.01183 [astro-ph.CO]} \BibitemShut {NoStop}%
\bibitem [{\citenamefont {Normann}\ and\ \citenamefont {Brevik}(2021)}]{Normann:2021bjy}%
  \BibitemOpen
  \bibfield  {author} {\bibinfo {author} {\bibfnamefont {B.~D.}\ \bibnamefont {Normann}}\ and\ \bibinfo {author} {\bibfnamefont {I.~H.}\ \bibnamefont {Brevik}},\ }\href {https://doi.org/10.1142/S0217732321501984} {\bibfield  {journal} {\bibinfo  {journal} {Mod. Phys. Lett. A}\ }\textbf {\bibinfo {volume} {36}},\ \bibinfo {pages} {2150198} (\bibinfo {year} {2021})},\ \Eprint {https://arxiv.org/abs/2107.13533} {arXiv:2107.13533 [gr-qc]} \BibitemShut {NoStop}%
\bibitem [{\citenamefont {Cruz}\ \emph {et~al.}(2020)\citenamefont {Cruz}, \citenamefont {Gonz\'alez},\ and\ \citenamefont {Palma}}]{Cruz:2018psw}%
  \BibitemOpen
  \bibfield  {author} {\bibinfo {author} {\bibfnamefont {N.}~\bibnamefont {Cruz}}, \bibinfo {author} {\bibfnamefont {E.}~\bibnamefont {Gonz\'alez}},\ and\ \bibinfo {author} {\bibfnamefont {G.}~\bibnamefont {Palma}},\ }\href {https://doi.org/10.1007/s10714-020-02712-z} {\bibfield  {journal} {\bibinfo  {journal} {Gen. Rel. Grav.}\ }\textbf {\bibinfo {volume} {52}},\ \bibinfo {pages} {62} (\bibinfo {year} {2020})},\ \Eprint {https://arxiv.org/abs/1812.05009} {arXiv:1812.05009 [gr-qc]} \BibitemShut {NoStop}%
\bibitem [{\citenamefont {Cruz}\ \emph {et~al.}(2021)\citenamefont {Cruz}, \citenamefont {Gonz\'alez},\ and\ \citenamefont {Palma}}]{Cruz:2019uya}%
  \BibitemOpen
  \bibfield  {author} {\bibinfo {author} {\bibfnamefont {N.}~\bibnamefont {Cruz}}, \bibinfo {author} {\bibfnamefont {E.}~\bibnamefont {Gonz\'alez}},\ and\ \bibinfo {author} {\bibfnamefont {G.}~\bibnamefont {Palma}},\ }\href {https://doi.org/10.1142/S0217732321500322} {\bibfield  {journal} {\bibinfo  {journal} {Mod. Phys. Lett. A}\ }\textbf {\bibinfo {volume} {36}},\ \bibinfo {pages} {2150032} (\bibinfo {year} {2021})},\ \Eprint {https://arxiv.org/abs/1906.04570} {arXiv:1906.04570 [gr-qc]} \BibitemShut {NoStop}%
\bibitem [{\citenamefont {Maartens}(1995)}]{Maartens:1995wt}%
  \BibitemOpen
  \bibfield  {author} {\bibinfo {author} {\bibfnamefont {R.}~\bibnamefont {Maartens}},\ }\href {https://doi.org/10.1088/0264-9381/12/6/011} {\bibfield  {journal} {\bibinfo  {journal} {Class. Quant. Grav.}\ }\textbf {\bibinfo {volume} {12}},\ \bibinfo {pages} {1455} (\bibinfo {year} {1995})}\BibitemShut {NoStop}%
\bibitem [{\citenamefont {Maartens}\ and\ \citenamefont {Mendez}(1997)}]{Maartens:1996dk}%
  \BibitemOpen
  \bibfield  {author} {\bibinfo {author} {\bibfnamefont {R.}~\bibnamefont {Maartens}}\ and\ \bibinfo {author} {\bibfnamefont {V.}~\bibnamefont {Mendez}},\ }\href {https://doi.org/10.1103/PhysRevD.55.1937} {\bibfield  {journal} {\bibinfo  {journal} {Phys. Rev. D}\ }\textbf {\bibinfo {volume} {55}},\ \bibinfo {pages} {1937} (\bibinfo {year} {1997})},\ \Eprint {https://arxiv.org/abs/astro-ph/9611205} {arXiv:astro-ph/9611205} \BibitemShut {NoStop}%
\bibitem [{\citenamefont {Eckart}(1940)}]{Eckart:1940te}%
  \BibitemOpen
  \bibfield  {author} {\bibinfo {author} {\bibfnamefont {C.}~\bibnamefont {Eckart}},\ }\href {https://doi.org/10.1103/PhysRev.58.919} {\bibfield  {journal} {\bibinfo  {journal} {Phys. Rev.}\ }\textbf {\bibinfo {volume} {58}},\ \bibinfo {pages} {919} (\bibinfo {year} {1940})}\BibitemShut {NoStop}%
\bibitem [{\citenamefont {Cruz}\ \emph {et~al.}(2017{\natexlab{c}})\citenamefont {Cruz}, \citenamefont {Cruz},\ and\ \citenamefont {Lepe}}]{Cruz:2017bcv}%
  \BibitemOpen
  \bibfield  {author} {\bibinfo {author} {\bibfnamefont {M.}~\bibnamefont {Cruz}}, \bibinfo {author} {\bibfnamefont {N.}~\bibnamefont {Cruz}},\ and\ \bibinfo {author} {\bibfnamefont {S.}~\bibnamefont {Lepe}},\ }\href {https://doi.org/10.1103/PhysRevD.96.124020} {\bibfield  {journal} {\bibinfo  {journal} {Phys. Rev. D}\ }\textbf {\bibinfo {volume} {96}},\ \bibinfo {pages} {124020} (\bibinfo {year} {2017}{\natexlab{c}})},\ \Eprint {https://arxiv.org/abs/1710.02607} {arXiv:1710.02607 [gr-qc]} \BibitemShut {NoStop}%
\bibitem [{\citenamefont {Cruz}\ \emph {et~al.}(2018{\natexlab{b}})\citenamefont {Cruz}, \citenamefont {Gonz\'alez}, \citenamefont {Lepe},\ and\ \citenamefont {S\'aez-Chill\'on~G\'omez}}]{Cruz:2018yrr}%
  \BibitemOpen
  \bibfield  {author} {\bibinfo {author} {\bibfnamefont {N.}~\bibnamefont {Cruz}}, \bibinfo {author} {\bibfnamefont {E.}~\bibnamefont {Gonz\'alez}}, \bibinfo {author} {\bibfnamefont {S.}~\bibnamefont {Lepe}},\ and\ \bibinfo {author} {\bibfnamefont {D.}~\bibnamefont {S\'aez-Chill\'on~G\'omez}},\ }\href {https://doi.org/10.1088/1475-7516/2018/12/017} {\bibfield  {journal} {\bibinfo  {journal} {JCAP}\ }\textbf {\bibinfo {volume} {12}},\ \bibinfo {pages} {017}},\ \Eprint {https://arxiv.org/abs/1807.10729} {arXiv:1807.10729 [gr-qc]} \BibitemShut {NoStop}%
\bibitem [{\citenamefont {Mohan}\ \emph {et~al.}(2017)\citenamefont {Mohan}, \citenamefont {Sasidharan},\ and\ \citenamefont {Mathew}}]{Mohan:2017poq}%
  \BibitemOpen
  \bibfield  {author} {\bibinfo {author} {\bibfnamefont {N.~D.~J.}\ \bibnamefont {Mohan}}, \bibinfo {author} {\bibfnamefont {A.}~\bibnamefont {Sasidharan}},\ and\ \bibinfo {author} {\bibfnamefont {T.~K.}\ \bibnamefont {Mathew}},\ }\href {https://doi.org/10.1140/epjc/s10052-017-5428-y} {\bibfield  {journal} {\bibinfo  {journal} {Eur. Phys. J. C}\ }\textbf {\bibinfo {volume} {77}},\ \bibinfo {pages} {849} (\bibinfo {year} {2017})},\ \Eprint {https://arxiv.org/abs/1708.02437} {arXiv:1708.02437 [gr-qc]} \BibitemShut {NoStop}%
\bibitem [{\citenamefont {Chimento}\ \emph {et~al.}(1997)\citenamefont {Chimento}, \citenamefont {Jakubi}, \citenamefont {Mendez},\ and\ \citenamefont {Maartens}}]{Chimento:1997ga}%
  \BibitemOpen
  \bibfield  {author} {\bibinfo {author} {\bibfnamefont {L.~P.}\ \bibnamefont {Chimento}}, \bibinfo {author} {\bibfnamefont {A.~S.}\ \bibnamefont {Jakubi}}, \bibinfo {author} {\bibfnamefont {V.}~\bibnamefont {Mendez}},\ and\ \bibinfo {author} {\bibfnamefont {R.}~\bibnamefont {Maartens}},\ }\href {https://doi.org/10.1088/0264-9381/14/12/019} {\bibfield  {journal} {\bibinfo  {journal} {Class. Quant. Grav.}\ }\textbf {\bibinfo {volume} {14}},\ \bibinfo {pages} {3363} (\bibinfo {year} {1997})},\ \Eprint {https://arxiv.org/abs/gr-qc/9710029} {arXiv:gr-qc/9710029} \BibitemShut {NoStop}%
\bibitem [{\citenamefont {Goodman}\ and\ \citenamefont {Weare}(2010)}]{Goodman_Ensemble_2010}%
  \BibitemOpen
  \bibfield  {author} {\bibinfo {author} {\bibfnamefont {J.}~\bibnamefont {Goodman}}\ and\ \bibinfo {author} {\bibfnamefont {J.}~\bibnamefont {Weare}},\ }\href@noop {} {\bibfield  {journal} {\bibinfo  {journal} {Communications in applied mathematics and computational science}\ }\textbf {\bibinfo {volume} {5}},\ \bibinfo {pages} {65} (\bibinfo {year} {2010})}\BibitemShut {NoStop}%
\bibitem [{\citenamefont {Foreman-Mackey}\ \emph {et~al.}(2013)\citenamefont {Foreman-Mackey}, \citenamefont {Hogg}, \citenamefont {Lang},\ and\ \citenamefont {Goodman}}]{Foreman-Mackey:2012any}%
  \BibitemOpen
  \bibfield  {author} {\bibinfo {author} {\bibfnamefont {D.}~\bibnamefont {Foreman-Mackey}}, \bibinfo {author} {\bibfnamefont {D.~W.}\ \bibnamefont {Hogg}}, \bibinfo {author} {\bibfnamefont {D.}~\bibnamefont {Lang}},\ and\ \bibinfo {author} {\bibfnamefont {J.}~\bibnamefont {Goodman}},\ }\href {https://doi.org/10.1086/670067} {\bibfield  {journal} {\bibinfo  {journal} {Publ. Astron. Soc. Pac.}\ }\textbf {\bibinfo {volume} {125}},\ \bibinfo {pages} {306} (\bibinfo {year} {2013})},\ \Eprint {https://arxiv.org/abs/1202.3665} {arXiv:1202.3665 [astro-ph.IM]} \BibitemShut {NoStop}%
\bibitem [{\citenamefont {Brout}\ \emph {et~al.}(2022)\citenamefont {Brout} \emph {et~al.}}]{Brout:2022vxf}%
  \BibitemOpen
  \bibfield  {author} {\bibinfo {author} {\bibfnamefont {D.}~\bibnamefont {Brout}} \emph {et~al.},\ }\href {https://doi.org/10.3847/1538-4357/ac8e04} {\bibfield  {journal} {\bibinfo  {journal} {Astrophys. J.}\ }\textbf {\bibinfo {volume} {938}},\ \bibinfo {pages} {110} (\bibinfo {year} {2022})},\ \Eprint {https://arxiv.org/abs/2202.04077} {arXiv:2202.04077 [astro-ph.CO]} \BibitemShut {NoStop}%
\bibitem [{\citenamefont {Tripp}(1998)}]{Tripp:1997wt}%
  \BibitemOpen
  \bibfield  {author} {\bibinfo {author} {\bibfnamefont {R.}~\bibnamefont {Tripp}},\ }\href@noop {} {\bibfield  {journal} {\bibinfo  {journal} {Astron. Astrophys.}\ }\textbf {\bibinfo {volume} {331}},\ \bibinfo {pages} {815} (\bibinfo {year} {1998})}\BibitemShut {NoStop}%
\bibitem [{\citenamefont {Kessler}\ and\ \citenamefont {Scolnic}(2017)}]{Kessler:2016uwi}%
  \BibitemOpen
  \bibfield  {author} {\bibinfo {author} {\bibfnamefont {R.}~\bibnamefont {Kessler}}\ and\ \bibinfo {author} {\bibfnamefont {D.}~\bibnamefont {Scolnic}},\ }\href {https://doi.org/10.3847/1538-4357/836/1/56} {\bibfield  {journal} {\bibinfo  {journal} {Astrophys. J.}\ }\textbf {\bibinfo {volume} {836}},\ \bibinfo {pages} {56} (\bibinfo {year} {2017})},\ \Eprint {https://arxiv.org/abs/1610.04677} {arXiv:1610.04677 [astro-ph.CO]} \BibitemShut {NoStop}%
\bibitem [{\citenamefont {Treu}\ and\ \citenamefont {Marshall}(2016)}]{Treu:2016ljm}%
  \BibitemOpen
  \bibfield  {author} {\bibinfo {author} {\bibfnamefont {T.}~\bibnamefont {Treu}}\ and\ \bibinfo {author} {\bibfnamefont {P.~J.}\ \bibnamefont {Marshall}},\ }\href {https://doi.org/10.1007/s00159-016-0096-8} {\bibfield  {journal} {\bibinfo  {journal} {Astron. Astrophys. Rev.}\ }\textbf {\bibinfo {volume} {24}},\ \bibinfo {pages} {11} (\bibinfo {year} {2016})},\ \Eprint {https://arxiv.org/abs/1605.05333} {arXiv:1605.05333 [astro-ph.CO]} \BibitemShut {NoStop}%
\bibitem [{\citenamefont {Jee}\ \emph {et~al.}(2019)\citenamefont {Jee}, \citenamefont {Suyu}, \citenamefont {Komatsu}, \citenamefont {Fassnacht}, \citenamefont {Hilbert},\ and\ \citenamefont {Koopmans}}]{Jee:2019hah}%
  \BibitemOpen
  \bibfield  {author} {\bibinfo {author} {\bibfnamefont {I.}~\bibnamefont {Jee}}, \bibinfo {author} {\bibfnamefont {S.~H.}\ \bibnamefont {Suyu}}, \bibinfo {author} {\bibfnamefont {E.}~\bibnamefont {Komatsu}}, \bibinfo {author} {\bibfnamefont {C.~D.}\ \bibnamefont {Fassnacht}}, \bibinfo {author} {\bibfnamefont {S.}~\bibnamefont {Hilbert}},\ and\ \bibinfo {author} {\bibfnamefont {L.~V.~E.}\ \bibnamefont {Koopmans}},\ }\href {https://doi.org/10.1126/science.aat7371} {\bibfield  {journal} {\bibinfo  {journal} {Science}\ }\textbf {\bibinfo {volume} {365}},\ \bibinfo {pages} {1134–1138} (\bibinfo {year} {2019})},\ \Eprint {https://arxiv.org/abs/1909.06712} {arXiv:1909.06712 [astro-ph.CO]} \BibitemShut {NoStop}%
\bibitem [{\citenamefont {Birrer}\ \emph {et~al.}(2019)\citenamefont {Birrer} \emph {et~al.}}]{Birrer:2018vtm}%
  \BibitemOpen
  \bibfield  {author} {\bibinfo {author} {\bibfnamefont {S.}~\bibnamefont {Birrer}} \emph {et~al.},\ }\href {https://doi.org/10.1093/mnras/stz200} {\bibfield  {journal} {\bibinfo  {journal} {Mon. Not. Roy. Astron. Soc.}\ }\textbf {\bibinfo {volume} {484}},\ \bibinfo {pages} {4726} (\bibinfo {year} {2019})},\ \Eprint {https://arxiv.org/abs/1809.01274} {arXiv:1809.01274 [astro-ph.CO]} \BibitemShut {NoStop}%
\bibitem [{\citenamefont {Rusu}\ \emph {et~al.}(2020)\citenamefont {Rusu} \emph {et~al.}}]{Rusu:2019xrq}%
  \BibitemOpen
  \bibfield  {author} {\bibinfo {author} {\bibfnamefont {C.~E.}\ \bibnamefont {Rusu}} \emph {et~al.},\ }\href {https://doi.org/10.1093/mnras/stz3451} {\bibfield  {journal} {\bibinfo  {journal} {Mon. Not. Roy. Astron. Soc.}\ }\textbf {\bibinfo {volume} {498}},\ \bibinfo {pages} {1440} (\bibinfo {year} {2020})},\ \Eprint {https://arxiv.org/abs/1905.09338} {arXiv:1905.09338 [astro-ph.CO]} \BibitemShut {NoStop}%
\bibitem [{\citenamefont {Chen}\ \emph {et~al.}(2019)\citenamefont {Chen} \emph {et~al.}}]{Chen:2019ejq}%
  \BibitemOpen
  \bibfield  {author} {\bibinfo {author} {\bibfnamefont {G.~C.~F.}\ \bibnamefont {Chen}} \emph {et~al.},\ }\href {https://doi.org/10.1093/mnras/stz2547} {\bibfield  {journal} {\bibinfo  {journal} {Mon. Not. Roy. Astron. Soc.}\ }\textbf {\bibinfo {volume} {490}},\ \bibinfo {pages} {1743} (\bibinfo {year} {2019})},\ \Eprint {https://arxiv.org/abs/1907.02533} {arXiv:1907.02533 [astro-ph.CO]} \BibitemShut {NoStop}%
\bibitem [{\citenamefont {Escamilla-Rivera}\ and\ \citenamefont {Torres~Castillejos}(2023)}]{Escamilla-Rivera:2022mkc}%
  \BibitemOpen
  \bibfield  {author} {\bibinfo {author} {\bibfnamefont {C.}~\bibnamefont {Escamilla-Rivera}}\ and\ \bibinfo {author} {\bibfnamefont {R.}~\bibnamefont {Torres~Castillejos}},\ }\href {https://doi.org/10.3390/universe9010014} {\bibfield  {journal} {\bibinfo  {journal} {Universe}\ }\textbf {\bibinfo {volume} {9}},\ \bibinfo {pages} {14} (\bibinfo {year} {2023})},\ \Eprint {https://arxiv.org/abs/2301.00490} {arXiv:2301.00490 [astro-ph.CO]} \BibitemShut {NoStop}%
\bibitem [{\citenamefont {Akiyama}\ \emph {et~al.}(2019)\citenamefont {Akiyama} \emph {et~al.}}]{EventHorizonTelescope:2019dse}%
  \BibitemOpen
  \bibfield  {author} {\bibinfo {author} {\bibfnamefont {K.}~\bibnamefont {Akiyama}} \emph {et~al.} (\bibinfo {collaboration} {Event Horizon Telescope}),\ }\href {https://doi.org/10.3847/2041-8213/ab0ec7} {\bibfield  {journal} {\bibinfo  {journal} {Astrophys. J. Lett.}\ }\textbf {\bibinfo {volume} {875}},\ \bibinfo {pages} {L1} (\bibinfo {year} {2019})},\ \Eprint {https://arxiv.org/abs/1906.11238} {arXiv:1906.11238 [astro-ph.GA]} \BibitemShut {NoStop}%
\bibitem [{\citenamefont {Akiyama}\ \emph {et~al.}(2022)\citenamefont {Akiyama} \emph {et~al.}}]{EventHorizonTelescope:2022wkp}%
  \BibitemOpen
  \bibfield  {author} {\bibinfo {author} {\bibfnamefont {K.}~\bibnamefont {Akiyama}} \emph {et~al.} (\bibinfo {collaboration} {Event Horizon Telescope}),\ }\href {https://doi.org/10.3847/2041-8213/ac6674} {\bibfield  {journal} {\bibinfo  {journal} {Astrophys. J. Lett.}\ }\textbf {\bibinfo {volume} {930}},\ \bibinfo {pages} {L12} (\bibinfo {year} {2022})},\ \Eprint {https://arxiv.org/abs/2311.08680} {arXiv:2311.08680 [astro-ph.HE]} \BibitemShut {NoStop}%
\bibitem [{\citenamefont {Denicol}\ \emph {et~al.}(2008)\citenamefont {Denicol}, \citenamefont {Kodama}, \citenamefont {Koide},\ and\ \citenamefont {Mota}}]{Denicol:2008ha}%
  \BibitemOpen
  \bibfield  {author} {\bibinfo {author} {\bibfnamefont {G.~S.}\ \bibnamefont {Denicol}}, \bibinfo {author} {\bibfnamefont {T.}~\bibnamefont {Kodama}}, \bibinfo {author} {\bibfnamefont {T.}~\bibnamefont {Koide}},\ and\ \bibinfo {author} {\bibfnamefont {P.}~\bibnamefont {Mota}},\ }\href {https://doi.org/10.1088/0954-3899/35/11/115102} {\bibfield  {journal} {\bibinfo  {journal} {J. Phys. G}\ }\textbf {\bibinfo {volume} {35}},\ \bibinfo {pages} {115102} (\bibinfo {year} {2008})},\ \Eprint {https://arxiv.org/abs/0807.3120} {arXiv:0807.3120 [hep-ph]} \BibitemShut {NoStop}%
\bibitem [{\citenamefont {Pu}\ \emph {et~al.}(2010)\citenamefont {Pu}, \citenamefont {Koide},\ and\ \citenamefont {Rischke}}]{Pu:2009fj}%
  \BibitemOpen
  \bibfield  {author} {\bibinfo {author} {\bibfnamefont {S.}~\bibnamefont {Pu}}, \bibinfo {author} {\bibfnamefont {T.}~\bibnamefont {Koide}},\ and\ \bibinfo {author} {\bibfnamefont {D.~H.}\ \bibnamefont {Rischke}},\ }\href {https://doi.org/10.1103/PhysRevD.81.114039} {\bibfield  {journal} {\bibinfo  {journal} {Phys. Rev. D}\ }\textbf {\bibinfo {volume} {81}},\ \bibinfo {pages} {114039} (\bibinfo {year} {2010})},\ \Eprint {https://arxiv.org/abs/0907.3906} {arXiv:0907.3906 [hep-ph]} \BibitemShut {NoStop}%
\bibitem [{\citenamefont {Romatschke}(2010)}]{Romatschke:2009im}%
  \BibitemOpen
  \bibfield  {author} {\bibinfo {author} {\bibfnamefont {P.}~\bibnamefont {Romatschke}},\ }\href {https://doi.org/10.1142/S0218301310014613} {\bibfield  {journal} {\bibinfo  {journal} {Int. J. Mod. Phys. E}\ }\textbf {\bibinfo {volume} {19}},\ \bibinfo {pages} {1} (\bibinfo {year} {2010})},\ \Eprint {https://arxiv.org/abs/0902.3663} {arXiv:0902.3663 [hep-ph]} \BibitemShut {NoStop}%
\bibitem [{\citenamefont {Reula}\ and\ \citenamefont {Nagy}(1997)}]{Reula:1995wk}%
  \BibitemOpen
  \bibfield  {author} {\bibinfo {author} {\bibfnamefont {O.~A.}\ \bibnamefont {Reula}}\ and\ \bibinfo {author} {\bibfnamefont {G.~B.}\ \bibnamefont {Nagy}},\ }\href {https://doi.org/10.1088/0305-4470/30/5/030} {\bibfield  {journal} {\bibinfo  {journal} {J. Phys. A}\ }\textbf {\bibinfo {volume} {30}},\ \bibinfo {pages} {1695} (\bibinfo {year} {1997})},\ \Eprint {https://arxiv.org/abs/gr-qc/9510068} {arXiv:gr-qc/9510068} \BibitemShut {NoStop}%
\end{thebibliography}%
\end{document}